\documentclass[aps,prd,reprint,groupedaddress,showpacs,nofootinbib]{revtex4-1}
\usepackage{hyperref}
\usepackage{slashed}
\usepackage{amsmath}
\usepackage{amssymb}
\usepackage{mathrsfs}
\usepackage{mathtools}
\usepackage{bbm}

\def\bra#1{\mathinner{\langle{#1}|}} 
\def\ket#1{\mathinner{|{#1}\rangle}} 
\newcommand{\braket}[2]{\langle #1|#2\rangle} 
\DeclarePairedDelimiter\abs{\lvert}{\rvert}
\DeclarePairedDelimiter\norm{\lVert}{\rVert}
\makeatletter
\let\oldabs\abs
\def\abs{\@ifstar{\oldabs}{\oldabs*}}
\let\oldnorm\norm
\def\norm{\@ifstar{\oldnorm}{\oldnorm*}}
\makeatother

\begin{document}

\title{Non-perturbative quantization of the electroweak model's electrodynamic sector}

\author{M. P. Fry}

\affiliation{University of Dublin, Trinity College, Dublin 2, Ireland}

\date{\today}

\begin{abstract}
Consider the Euclidean functional integral representation of any physical process in the electroweak model.
Integrating out the fermion degrees of freedom introduces twenty-four fermion determinants.
These multiply the Gaussian functional measures of the Maxwell, $Z$, $W$ and Higgs fields to give an effective functional measure.
Suppose the functional integral over the Maxwell field is attempted first.
This paper is concerned with the large amplitude behavior of the Maxwell effective measure.
It is assumed that the large amplitude variation of this measure is insensitive to the presence of the $Z$, $W$ and $H$ fields; they are assumed to be a subdominant perturbation of the large amplitude Maxwell sector.
Accordingly, we need only examine the large amplitude variation of a single QED fermion determinant.
To facilitate this the Schwinger proper time representation of this determinant is decomposed into a sum of three terms.
The advantage of this is that the separate terms can be non-perturbatively estimated for a measurable class of large amplitude random fields in four dimensions.
It is found that the QED fermion determinant grows faster than $\exp \left[ce^2\int\mathrm d^4x\, F_{\mu\nu}^2\right]$, $c>0$, in the absence of zero mode supporting random background potentials.
This raises doubt on whether the QED fermion determinant is integrable with any Gaussian measure whose support does not include zero mode supporting potentials.

Including zero mode supporting background potentials can result in a decaying exponential growth of the fermion determinant.
This is \textit{prima facie} evidence that Maxwellian zero modes are necessary for the non-perturbative quantization of QED and, by implication, for the non-perturbative quantization of the electroweak model.
\end{abstract}


\maketitle

\section{Introduction}
  It is not known if the electroweak model can be non-perturbatively  quantized.
 This requires the convergence of the unexpanded functional  integrals over all classical field configurations for the vacuum  expectation values of its field operators.
 It is assumed  that the integrals have been continued to Euclidean space to  make mathematical sense out of them and that ultraviolet and volume  cutoffs are in place in their integrands.
 Their introduction will be  discussed later.
 Since the quantization is non-perturbative most  of the functional integrals cannot be done explicitly.
 Therefore,  the criteria for the non-perturbative renormalization of the  model are not known \textit{ab initio}.
 Immediately one is confronted  with an external field problem: do the regulated integrands grow slowly  enough with large amplitude field variations for the functional  integrals to converge? It is the aim of this paper to examine  this minimal requirement for the non-perturbative quantization  of the electroweak model.

       Presumably the order of doing the functional  integrals is irrelevant aside from their technical difficulty.
  If so, it is reasonable to begin with what is well-known.
  Accordingly, we first integrate out the fermions.
 Then the answer  to the above question partly depends on knowing the strong field  behavior of each of the  6 lepton and 3$\times$6 quark determinants obtained by integrating out  the three generations of leptons and quarks, including their three  colors.
 For example, the electron and its associated  neutrino field\footnote{The extension of the model to massive neutrinos and their  mixing is not considered here as it will not affect the main results  of this paper.} contribute the following factor to the Euclidean functional integral representation of any electroweak process after spontaneous symmetry breaking:
 \begin{widetext}
     \begin{align}
         \begin{split}
             \det &\left[ \slashed P + m_e + e\slashed A 
                 + \frac{g}{2\cos\theta_W}\slashed Z \left( \frac{1-\gamma_5}{2}\right)
                 - \frac{g\sin^2\theta_W}{\cos\theta_W}\slashed Z
                 + \frac{gm_e}{2M_W}H\right]\times\\
             &\det \left[ \slashed P - \frac{g}{2\cos\theta_W}\slashed Z \left (
                 \frac{1-\gamma_5}{2}\right )
                 -\frac{g^2}{2}\slashed W^+
                  \left ( \frac{1-\gamma_5}{2}\right ) S_e\slashed W^-
             \left ( \frac{1-\gamma_5}{2}\right )\right ].
         \end{split}
         \tag{1.1}
         \label{eq:onepointone}
     \end{align}
 \end{widetext}
  
Here $A_\mu$, $Z_\mu$, $W^\pm_\mu$, and $H$ are the Maxwell, neutral and charged vector boson and Higgs fields; $S_e$, the inverse of the operator in brackets in the first determinant, is the electron propagator in the presence  of the $A$, $Z$ and $H$ fields; $m_e$ and $M_W$  are the electron and $W$-boson  masses; $e$ is the positron electric charge; $\theta_W$ is the Weinberg angle and $g = e/\sin\theta_W$.
The result in (\ref{eq:onepointone}) follows by inspection of the  electroweak Lagrangian \cite{1} and an elementary integration over the  electroweak action quadratic in the fermion fields \cite{2}.
The twenty-four  determinants multiply the Gaussian measures  $d\mu(A)\,d\mu(Z)\,d\mu(W)\,d\mu(H)$ as does the remainder of the electroweak  action denoted by $\exp\left[-\int \mathrm d^4x\, \mathcal L (A, Z, W^\pm, H)\right]$. Considering the complexity of the Feynman rules in the 't Hooft-Feynman gauge a non-perturbative calculation may simplify in the unitary gauge. The absence of the Goldstone bosons $\chi, \varphi^{\pm}$ in the determinants in (\ref{eq:onepointone}) indicates that this gauge has been selected.

An ultraviolet cutoff has to be introduced into the $A, Z, W$ and $H$ field propagators. As these fields are to be integrated over they are assumed to be tempered distributions. In order to calculate the fermion determinants these fields need to be smoothed following the procedure outlined at the beginning of Sec. VII for QED. The smoothing procedure introduces an ultraviolet cutoff in the associated propagators when calculating the fields' covariances with the above Gaussian gauge-fixed measures as in Eq.(\ref{eq:7.2}). Thus the ultraviolet cutoffs are introduced by functionally integrating the electroweak model.

The fermion determinants contain all fermion loops and hence the anomalies. The process for cancelling them in this paper begins by noting that the determinants, such as those in (\ref{eq:onepointone}), are ill-defined as they stand. Mathematical sense can be made of them by subtracting out all loops whose degree of divergence is 2, 1 and 0. The subtraction process is illustrated by (\ref{eq:F1}) in Appendix F for the case of QED. As a representative example consider the  
$\gamma W^+ W^-$ triangle graph containing three fermion propagators. Schematically the electron neutrino determinant in (\ref{eq:onepointone}) is subtracted so that
det$\rightarrow \exp[\Pi(ee\nu_e)+\text{other subtractions}]\times \text{det}_R $, where det$_R$ is a well-defined remainder determinant similar to det$_5$  in (\ref{eq:F1}) and (\ref{eq:F2}); $\Pi(ee\nu_e)$ denotes the first generation lepton triangle graph for $\gamma\rightarrow W^+ W^-$. When the 23 remaining determinants are subtracted the exponentiated subtractions combine to give the following result for the sum of all the graphs contributing to the first generation $\gamma W^+ W^-$  triangle anomaly:
 \begin{align}
 \begin{split}
  \exp \{ \Pi(ee\nu_e) &+ 3[\Pi(ddu) + \Pi(uud)] |V_{ud}|^2 \\
                  &+   3[\Pi(ssu) + \Pi(uus)] |V_{us}|^2 \\
                  &+   3[\Pi(bbu) + \Pi(uub)] |V_{ub}|^2 \\
                  &+ \text{other subtractions} \} \times \Pi_{i=1}^{24}\text{det}_{R_i}.
 \label{eq:onepointtwo}
 \end{split}
 \tag{1.2}
 \end{align}
                          
Here $u, d, s ,b$ refer to quark flavors and $V_{ij}$   is the CKM quark mixing matrix \cite{1}. The anomaly is removed by subtracting out the zero-mass limit of these graphs which we denote by $\Pi_0$  . Then the anomaly bearing graphs reduce to

\begin{equation}
\exp\{\Pi_0(ee\nu_e)+3[\Pi_0(uud)+\Pi_0(ddu)](|V_{ud}|^2+|V_{us}|^2+|V_{ub}|^2)\}
\tag{1.3}
\label{eq:onepointthree}
\end{equation}
since there is no difference between the free u,d,s and b propagators in the massless limit. Noting that the unitarity of the CKM matrix requires the sum of the matrix elements in (\ref{eq:onepointthree}) to be one, the sum of the color weighted $\gamma$ -vertices in (\ref{eq:onepointthree}) results in the cancellation of the first generation $\gamma W^+ W^-$ triangle anomaly. This procedure can be continued until all of the three and four leg anomalies in the three generations cancel as they are known to do. These determinant regularizations should be done before they are inserted into the functional integrals over the gauge and Higgs fields.

Summarizing, it is necessary to define the fermion determinants by removing their ill-defined loops by making subtractions that are then either renormalized or cancelled among themselves. This happens to lead to anomaly cancellation at the three and four external leg level. Of course it has not been proved that the product of the remainder determinants is free of terms that can block the non-perturbative renormalization of the electroweak model \cite{55}.

It is known that when $\Pi_{i=1}^{24}\text{det}_{R_i}$ is loop-expanded it contains an exponentiated sum of absolutely convergent graphs beginning with the pentagon graph. These can be calculated in a manifestly gauge invariant way and cannot contain anomalies. The fact that the perturbative expansion of $\Pi_{i=1}^{24}\text{det}_{R_i}$ is anomaly-free leaves open the possibility that this determinant product may eventually be shown to be part of a non-perturbative, anomaly-free, gauge preserving regularization of the electroweak model.

Assuming the functional integrals converge the process of renormalization follows next with the introduction of counterterms to remove the regulators. Presumably the result is in terms of the physical parameters $e$, $M_W$, $M_Z$, $M_H$, $m_i$ -the charged fermion masses- and the renormalized quark mixing matrix $V_{ij}$ after continuing from an intermediate renormalization scheme in Euclidean space to on-shell renormalization in Minkowski space.

The observation that $\mathcal L$  is no more than quadratic in $A_\mu$,  that $A_\mu$ does not couple directly to $H$, that a considerable amount  is known about the QED determinant  $\det(\slashed P -e\slashed A + m)$, and that the  regularization of the electrodynamic sector is straightforward  suggests that the next simplest functional integration should be  over the Maxwell field.
 Supppose this is decided.
Twenty-one   of the twenty-four fermion determinants involve  the Maxwell field as it appears in the electron's  determinant in (\ref{eq:onepointone}) with different charges. Should their combined large amplitude $A$-field variation increase faster than $\exp\left[ce^2 \int \mathrm d^4 x\, F_{\mu\nu}^2  \right]$, $c>0$
then the integration over the Maxwell field with any Gaussian measure  would be divergent, and the non-perturbative quantization of the  electroweak model would be doubtful.
The $F_{\mu\nu}$-dependence is expected since  the determinants are gauge invariant.

It is assumed that the strong Maxwell field behavior of these determinants can be obtained by decoupling them from the electroweak model by setting $g=0$. Future theorems dealing with the assumed sub-dominant growth of the remainder determinants can and should be produced. Noting this, there remains a product of twenty one
determinants of the form det$(\slashed{P} - q\slashed{A} + m)$ so that we need only calculate one of them. Accordingly, this paper considers the the non-perturbative quantization of the electroweak model's electrodynamic sector. It is found that this can be done only under restrictive conditions. If the sub-dominance of the remainder determinants assumed here is valid then these conditions extend to the complete electroweak model.

\section{Preliminaries}
 Confining attention to QED, sense has to be made of the infinite  dimensional determinant $\det( \slashed P - e\slashed A + m)$, where $e>0$ from here on.
 It  is first normalized to one when $e=0$ by dividing it by $\det( \slashed P + m)$ to  get $\det(1-eS\slashed A)$, where $S$ is the free electron propagator.
 To make this  well-defined it has to be regularized and made ultraviolet finite  by a second order charge renormalization subtraction.
 A representation  of the regulated and renormalized determinant, denoted by $\det_\mathrm{ren}$, is  given by Schwinger's proper time definition \cite{3}
 \begin{widetext}
     \begin{align}
         \ln\det{}_\mathrm{ren}(1-e_0S\slashed A)=\frac{1}{2}\int_0^\infty
         \frac{\mathrm d t}{t}
         \left( \mathrm{Tr}\left\{ e^{-P^2 t}-\exp\left[ -\left( D^2 + \frac{e_{\mathrm{o}}}{2}\sigma_{\mu\nu}F_{\mu\nu}\right)t\right]\right\}
         +\frac{e^2_{\mathrm{o}}\| F\|^2}{24\pi^2}\right)e^{-tm^2_{\mathrm{o}}},
         \tag{2.1}
         \label{eq:twopointone}
     \end{align}
 \end{widetext}
 where $D_\mu = P_\mu -e_\mathrm{o}A_\mu$, $\sigma_{\mu\nu}=[\gamma_\mu,\,\gamma_\nu]/2i$, $\gamma_\mu^\dagger=-\gamma_\mu$, $\|F\|^2=\int \mathrm d^4x\, F_{\mu\nu}^2$, and 
$e_{\mathrm{o}}$, $m_{\mathrm{o}}$ are the unrenormalized charge and mass. The last term in (\ref{eq:twopointone}) results in a second-order charge renormalization subtraction in the one-particle irreducible photon self-energy $\Pi(k^2)$ at zero momentum transfer as in Eq.(\ref{eq:C7}), Appendix C. Therefore, as long as $A_{\mu}$ remains a classical field $e_{\mathrm{o}}$ and $m_{\mathrm{o}}$ are the physical parameters e and m. Quantizing $A_{\mu}$ by integrating over it will require a further charge renormalization subtraction given by $1/e_{\mathrm{o}}^2=1/ {e^2}+\Pi(0,e_{\mathrm{o}}^2D_{\mathrm{o}})$, where $\Pi(0,e_{\mathrm{o}}^2D_{\mathrm{o}})$ is the 1PI photon self-energy at $k^2=0$ with the one-loop contribution omitted. It is a functional of the exact unrenormalized photon propagator $D_{\mathrm{o}}$ with $\Pi(0,0)=0$; it is made finite by the regularization procedure outlined in Sec. VII. As renormalization will not be considered further the subscript o will be dropped in (\ref{eq:twopointone}) 
with the understanding that e and m are the {\it{unrenormalized}} charge and mass in what follows.

 Having defined $\det_\mathrm{ren}$ the effective measure for the Maxwell field  integration is  
 \begin{align}
     \mathrm d\mu(A)=Z^{-1}\mathrm d\mu_0(A)\det{}_\mathrm{ren}(1-eS\slashed A)
     \tag{2.2}
     \label{eq:twopointtwo}
 \end{align}
 where the gauge-fixed Gaussian measure for the random potential $A_\mu$  is now denoted by $d\mu_0$.
 It has mean zero and covariance 
 \begin{align}
     \int \mathrm d\mu_0\, A_\mu(x)A_\nu(y)=D_{\mu\nu}(x-y),
     \tag{2.3}
     \label{eq:twopointthree}
 \end{align}
 where $D_{\mu\nu}$ is the photon propagator in a fixed gauge.
 The vacuum-vacuum  amplitude $Z$ in (\ref{eq:twopointthree}) is
 \begin{align}
     Z=\int \mathrm d\mu_0\,\det{}_\mathrm{ren},
     \tag{2.4}
     \label{eq:twopointfour}
 \end{align}
 so that $\int \mathrm d\mu(A)=1$.
 The measure (\ref{eq:twopointtwo}) appears in the non-perturbative  calculation of every physical process in QED such as the Euclidean  Green function for $2n$ external fermions and $m$ photons,
 \begin{widetext}
    \begin{align}
        \begin{split}
            S_{\mu_1\ldots\mu_m}&(x_1,\ldots,x_n;y_1,\ldots,y_n;z_1,\ldots,z_m)\\
            &=Z^{-1}\int \mathrm d\mu_0(A)\, \det{}_\mathrm{ren}(1-eS\slashed A)\det\left[S(x_i,y_j|eA)\right]_{i,j=1}^n\prod_{k=1}^mA_{\mu_k}(z_k),
        \end{split}
        \tag{2.5}
        \label{eq:twopointfive}
    \end{align}
 \end{widetext}
 where $S(x,y|eA)$ is the electron propagator in the external potential $A_\mu$.

Any attempt to calculate the integrals in (\ref{eq:twopointfour}) and (\ref{eq:twopointfive}) will  encounter ultraviolet divergences that require regularization.
 How  this regularization is introduced will be discussed in Sec.
 VII.
 In  addition $Z$ requires a volume cutoff that will be discussed in  in Sec.
 VII as well.
 A volume cutoff  enters QED solely by its determinant to render the vacuum energy  finite when the determinant is integrated.
 Assuming that the  functional integrations in (\ref{eq:twopointfour}) and (\ref{eq:twopointfive}) converge, there remains  the task of removing the ultraviolet regulator and volume cutoff  by some as yet unknown non-perturbative renormalization procedure  that preserves the unitarity of $S$-matrix elements.
 The difficulty of  implementing this procedure cannot be overstated.

 Whether the functional integrals in (\ref{eq:twopointfour}) and (\ref{eq:twopointfive}) converge  depends on $\det_\mathrm{ren}$'s behavior for large amplitude variations  of a measurable set of random fields $F_{\mu\nu}$ on $\mathbb R^4$.
 Since $e$ always  multiplies $F_{\mu\nu}$   it will be sufficient to consider the strong coupling  behavior of $\det_\mathrm{ren}$.

       This leads to one of the main results of this paper.
       Although (\ref{eq:twopointone})  is compact and intuitive it -- and all other representations -- have so far  failed to give any explicit information on the strong coupling  behavior of $\det_\mathrm{ren}$   for random fields on $\mathbb R^4$.
 To remedy this  an exact representation of $\ln\det_\mathrm{ren}$    is derived from (\ref{eq:twopointone})  that facilitates its strong coupling analysis.
 Noting that in  Euclidean space $F_{\mu\nu}$ may be regarded as a static, four-dimensional  magnetic field, the new representation breaks $\ln\det_\mathrm{ren}$    into a sum  of three terms that expose its competing magnetic properties, namely,
 \begin{align}
    \begin{split}
        \ln\det{}_\mathrm{ren}  &= \mbox{diamagnetism} + \mbox{paramagnetism} \\ &\quad + \mbox{charge renormalization}.
    \end{split}
    \tag{2.6}
    \label{eq:twopointsix}
 \end{align}

 The advantage of representation (\ref{eq:twopointsix}) of $\det{}_\mathrm{ren}$ is that the strong coupling  analysis of its separate terms is far easier than their combined form  in (\ref{eq:twopointone}).
 The derivation of (\ref{eq:twopointsix}) is given in Sec. III.
 Suffice it to say  here that the sum of the diamagnetic term (Sec.IV) and charge  renormalization term (Sec.VI) contribute to $\det_\mathrm{ren}$'s strong coupling  growth while the paramagnetic term (Sec.V) slows it down.
 Therefore, the  non-perturbative quantization of QED critically depends on the  paramagnetic term and the class of background fields on which it depends.
 {\it Prima facie} evidence is given that zero mode supporting background fields are necessary  for the non-perturbative quantization of QED.
 The presence of  substantial numbers of zero modes in the lattice functional integration  of QED in its chirally broken phase has been noted \cite{4,5}.
 Our result  and this observation suggest that Maxwellian zero modes will play a  key role in deciding whether the electroweak model can be  non-perturbatively quantized.
 Our conclusions are summarized in  Secs.VIC and VIII, and the appendices deal with mathematical details.

\section{Representation of $\boldsymbol{\det_\mathrm{ren}}$}
The objective is to obtain an expression for $\det_\mathrm{ren}$ that manifests  the interplay of diamagnetism, paramagnetism and charge renormalization  in its strong coupling behavior for random, static, four-dimensional  magnetic fields.
Rewrite (\ref{eq:twopointone}) as
\begin{widetext}
    \begin{align}
        \begin{split}
            \ln&\,\det{}_\mathrm{ren}\\
            &=\frac{1}{2}\int_0^\infty \frac{\mathrm dt}{t}\,e^{-tm^2}
            \left[ 4\mathrm{Tr}\left( e^{-P^2t} - e^{-D^2t}\right)
                -\frac{e^2\|F\|^2}{48\pi^2}
                + \mathrm{Tr} \left( e^{-D^2t} -\exp\left[-\left(D^2+ \frac{e}{2}\sigma_{\mu\nu}F_{\mu\nu}\right)t\right] \right)
            + \frac{e^2\| F\| ^2}{16\pi^2}\right],
        \end{split}
        \tag{3.1}
        \label{eq:threepointone}
    \end{align}
\end{widetext}
where the trace over spin was made in the first term to give a factor  of 4.
Then (\ref{eq:threepointone}) becomes
\begin{widetext}
    \begin{align}
        \ln\det{}_\mathrm{ren}=2\ln\det{}_\mathrm{SQED}
        + \frac{1}{2}\int_0^\infty \frac{\mathrm dt}{t}\,e^{-tm^2}
        \left[ \mathrm{Tr}\left( e^{-D^2t}-\exp\left[-\left(D^2+ \frac{e}{2}\sigma_{\mu\nu}F_{\mu\nu}\right)t\right] \right)
            + \frac{e^2\| F\| ^2}{16\pi^2}\right],
            \tag{3.2}
        \label{eq:threepointtwo}
    \end{align}
\end{widetext}
where $\ln\det_\mathrm{SQED}$ is the proper time definition of the formal scalar QED determinant  $\ln\det\left\{\left[( P - e A)^2 + m^2 \right]/(P^2 + m^2 )\right\}$ with on-shell charge renormalization:
    \begin{align}
        \begin{split}
            \ln&\det{}_\mathrm{SQED}\\
            &=\int_0^\infty \frac{\mathrm dt}{t}
            \left[ \mathrm{Tr}\left( e^{-P^2t} - e^{-D^2t}\right)
                -\frac{e^2\|F\|^2}{192\pi^2}
            \right]e^{-tm^2},
        \end{split}
        \tag{3.3}
        \label{eq:threepointthree}
    \end{align}
 Alternatively, $\ln\det_\mathrm{SQED} = -S_\mathrm{SQED}$, where $S_\mathrm{SQED}$    is the one-loop effective  action of scalar QED.

 Now consider the remaining terms in (\ref{eq:threepointtwo}) and use the operator  identity 
      \begin{align}
          \begin{split}
              e^{-t\left(D^2+\frac{1}{2}e\sigma F\right)} &- e^{-tD^2}\\
              &= -\int_0^t\mathrm d s\,e^{-(t-s)\left(D^2+\frac{1}{2}e\sigma F\right)}\frac{1}{2}e\sigma F e^{-sD^2}.
          \end{split}
          \tag{3.4}
          \label{eq:threepointfour}
      \end{align}
      
      A derivation of (\ref{eq:threepointfour}) is given in \cite{6}.
Iterating it twice gives 
\begin{widetext}
    \begin{align}
        \begin{split}
            &e^{-t\left(D^2+\frac{1}{2}e\sigma F\right)} - e^{-tD^2}\\
            = &-\int_0^t\mathrm d s\,e^{-(t-s)D^2}\frac{1}{2}e\sigma F e^{-sD^2}\\
              &+\int_0^t\mathrm d s_1\,\int_0^{t-s_1}\mathrm d s_2\,
            e^{-(t-s_1-s_2)D^2}\frac{1}{2}e\sigma F
            e^{-s_2D^2}\frac{1}{2}e\sigma F e^{-s_1D^2}\\
            &-\int_0^t\mathrm ds_1\,\int_0^{t-s_1}\mathrm ds_2\,\int_0^{t-s_1-s_2}\mathrm ds_3\,
            e^{-(t-s_1-s_2-s_3)\left(D^2+\frac{1}{2}e\sigma F\right)}
            \frac{1}{2}e\sigma F e^{-s_3D^2}
            \frac{1}{2}e\sigma F e^{-s_2D^2}
            \frac{1}{2}e\sigma F e^{-s_1D^2}.
        \end{split}
        \tag{3.5}
        \label{eq:threepointfive}
    \end{align}
\end{widetext}
      
      Define the determinant $\det_3$   by 
      \begin{widetext}
          \begin{align}
              \begin{split}
                  \ln\,\mbox{$\det_3$}\left(1+\Delta_A^{1/2} \frac{1}{2}e\sigma F\Delta_A^{1/2}\right)
                  &= \int_0^\infty \frac{\mathrm dt}{t}\,e^{-tm^2}\mathrm{Tr}\biggl(\int_0^t \mathrm d s_1\,\int_0^{t-s_1}\mathrm d s_2\,\int_0^{t-s_1-s_2}\mathrm d s_3\ \\
                   &\times e^{-(t-s_1-s_2-s_3)\left(D^2+\frac{1}{2}e\sigma F\right)}
                  \frac{1}{2}e\sigma F e^{-s_3D^2}
                  \frac{1}{2}e\sigma F e^{-s_2D^2}
                  \frac{1}{2}e\sigma F e^{-s_1D^2} \biggr),
                  \end{split}
                  \tag{3.6}
                  \label{eq:threepointsix}
          \end{align}
      \end{widetext}
 where $\Delta_A^{1/2} = (D^2  + m^2 )^{-1/2}$.
 Before proceeding with the derivation of (\ref{eq:twopointsix})  it is important to explain what the left-hand side of (\ref{eq:threepointsix}) means \cite{7,8,9,10,11}.

Thus $\det_3$  is the regularized determinant defined by 
\begin{align}
    \mbox{$\det_3$}(1+T)=\det\left[(1+T)\exp\left(-T+\frac{1}{2}T^2\right)\right],
    \tag{3.7}
    \label{eq:threepointseven}
\end{align}
provided $T\in \mathscr I_3$.
The trace ideal $\mathscr I_p$ ($1\le p< \infty$) is defined as those  compact operators $T$ with  $\|T\|_p^p   = \mathrm{Tr}((T^\dagger T)^{p/2})<\infty$ \cite{8,9,10}.
Because  $T$ is compact its eigenvalues are discrete and have finite multiplicity.
Therefore, the left-hand side of (\ref{eq:threepointsix}) requires that the operator $\Delta_A^{1/2}\sigma F\Delta_A^{1/2}\in\mathscr I_3$.
This is shown in Appendix A for $F_{\mu\nu}\in \cap_{p>2}L^p (\mathbb R^4)$ and $m \neq 0$.
Note that  this allows zero mode supporting potentials $A_\mu(x)$ with their necessary  $1/|x|$ fall off for $|x|\rightarrow\infty$.
The equivalence of the two sides of (\ref{eq:threepointsix})  follows from Theorem 7.2 in \cite{7} where an outline of its proof is given.
 Because of the inaccessibility of \cite{7} and the importance of $\det_3$   to this  paper a proof is given in Appendix B.
More will be said about $\det_3$   in  Sec.
V.
But already we anticipate that its presence in $\det_\mathrm{ren}$    will be a  calculational advantage as it deals with a self-adjoint operator acting  on countable, square-integrable eigenstates.
Put differently, $\det_3$'s  calculation reduces to a manageable quantum mechanical problem on bound  state energy levels as discussed in Sec.
VB.

Continuing with the derivation of (\ref{eq:twopointsix}), insert (\ref{eq:threepointfive}) and (\ref{eq:threepointsix})  in (\ref{eq:threepointtwo}) to obtain
      \begin{widetext}
          \begin{align}
              \begin{split}
                  \ln&\,\mbox{$\det{}_\mathrm{ren}$}=2\ln\det{}_\mathrm{SQED} + \frac{1}{2}\ln\mbox{$\det_3$}\left(1+\Delta_A^{1/2}\frac{1}{2}e\sigma F\Delta_A^{1/2}\right)\\
                  &+\frac{e^2}{8}\int_0^\infty \frac{\mathrm dt}{t}\,e^{-tm^2}
                  \left(\frac{1}{4\pi^2}\|F\|^2
                  -\mathrm{Tr}\int_0^t\mathrm ds_1\,\int_0^{t-s_1}\mathrm ds_2
                  e^{-(t-s_1-s_2)D^2}\sigma Fe^{-s_2D^2}\sigma Fe^{-s_1D^2}\right).
              \end{split}
              \tag{3.8}
              \label{eq:threepointeight}
          \end{align}
      \end{widetext}

      It is shown in Appendix C that the last term in (\ref{eq:threepointeight}) can be simplified  to give the promised three-term representation of $\ln\det_\mathrm{ren}$:
      \begin{widetext}
          \begin{align}
              \begin{split}
                  \ln\mbox{$\det{}_\mathrm{ren}$}
                  =2\ln\det{}_\mathrm{SQED} + \frac{1}{2}\ln\mbox{$\det_3$}\left(1+\Delta_A^{1/2}\frac{1}{2}e\sigma F\Delta_A^{1/2}\right)
                  +e^2\int_0^\infty \mathrm dt\,e^{-tm^2}
                  \left[\frac{1}{32\pi^2t}\|F\|^2
                      -\frac{1}{2} \mathrm{Tr}
                  \left(e^{-tD^2}F_{\mu\nu}\Delta_AF_{\mu\nu}\right)\right],
              \end{split}
              \tag{3.9}
              \label{eq:threepointnine}
          \end{align}
      \end{widetext}
      where $\Delta_A = (D^2 + m^2)^{-1}$.

Equation (\ref{eq:threepointnine}) is equivalent to (\ref{eq:twopointone}), and each term is separately  well-defined and gauge invariant.
Their order follows that in (\ref{eq:twopointsix}).
The signs of the first  two terms and their connection with diamagnetism and paramagnetism are  discussed in the following sections.
The last term is connected with  charge renormalization and is manifestly positive due to QED's lack of  asymptotic freedom.

\section{Strong coupling behavior of $\boldsymbol{\det_\mathrm{SQED}}$}
Let the amplitude of $F_{\mu\nu}(x)$ be set by the parameter $\mathscr{F}$ which has the  dimension of $L^{-2}$.
Then break the integral in (\ref{eq:threepointthree}) into $\int_0^{1/e\mathscr{F}}$ and $\int_{1/e\mathscr{F}}^\infty$ and use Kato's inequality in the form \cite{12,13,14,15}
\begin{align}
    \mathrm{Tr}\left(e^{-P^2t}-e^{-(P-eA)^2t}\right)\ge0,
    \tag{4.1}
    \label{eq:fourpointone}
\end{align}
to obtain
\begin{align}
    \begin{split}
        &\ln\mbox{$\det_\mathrm{SQED}$}\ge\\
        &\int_0^{1/e\mathscr{F}}\frac{\mathrm dt}{t}\,
        \left[ \mathrm{Tr}\left(e^{-P^2t}-e^{-(P-eA)^2t}\right) - \frac{e^2\|F\|^2}{192\pi^2}\right]
        e^{-tm^2}\\
        &\quad- \frac{e^2\|F\|^2}{192\pi^2}\int_{1/e\mathscr F}^\infty \frac{\mathrm dt}{t}\,
        e^{-tm^2}.
    \end{split}
    \tag{4.2}
    \label{eq:fourpointtwo}
\end{align}
The inequality in (\ref{eq:fourpointone}) reflects the diamagnetism of charged scalar  bosons: on average the energy levels of such bosons increase  in a magnetic field.
This explains the first term in (\ref{eq:twopointsix}).
The selection of $e\mathscr F$ as the scaling parameter is discussed below.

The first integral in (\ref{eq:fourpointtwo}) is dominated by its small-$t$  behavior for $e\gg1$.
Accordingly, make the heat kernel expansion
\begin{widetext}
    \begin{align}
       \begin{split}
        \mathrm{Tr}\left(e^{-P^2t}-e^{-(P-eA)^2t}\right)
        &=\frac{1}{16\pi^2}\int\mathrm d^4x\,
        \biggl[\frac{e^2}{12}F_{\mu\nu}^2
        +\frac{te^2}{120}F_{\mu\nu}\nabla^2F_{\mu\nu}\\
        &+\frac{t^2e^2}{1680}F_{\mu\nu}\nabla^4F_{\mu\nu}
        +\frac{t^2e^4}{1440}\left[(^\star F_{\mu\nu}F_{\mu\nu})^2
        -7(F_{\mu\nu}^2)^2\right] \biggr]+ \mathrm O(t^3),
        \end{split}
        \tag{4.3} 
        \label{eq:fourpointthree}
    \end{align}
\end{widetext}
where $^\star F_{\mu\nu}=\frac{1}{2}\epsilon_{\mu\nu\alpha\beta}F_{\alpha\beta}$.
The $\mathrm O(F^2)$ terms follow from the result for  $\ln\det_\mathrm{SQED}$ in (C6); the $\mathrm O(F^4)$ term is inferred from Schwinger's  constant field result for scalar QED \cite{3}

      To the author's knowledge there is no proof that QED heat kernel  expansions are asymptotic series in $t$ although this is generally  assumed.
      Referring to (\ref{eq:fourpointthree}) it is evident that continuing the  expansion in powers of $t$ requires that $F_{\mu\nu}$ be infinitely  differentiable ($C^\infty$).
So this is a necessary condition.
In Sec.VII  we will introduce an ultraviolet regulator by convoluting the  potential $A_\mu$ with a function of rapid decrease.
The resulting  smoothed potential is $C^\infty$.
Anticipating Sec. VII we will now assume  the fields in (\ref{eq:fourpointthree}) are $C^\infty$.
With this understanding the expansion  in (\ref{eq:fourpointthree}) will now be assumed to be asymptotic so that the  truncation error after $N$ terms is
\begin{align}
    \begin{split}
        &\mathrm{Tr}\left(e^{-P^2t}-e^{-(P-eA)^2t}\right)\\
        &\quad -\sum_{n=0}^Na_n(eF)t^n\underset{t\searrow0}{\sim}a_M(eF)t^M,
    \end{split}
        \tag{4.4}
    \label{eq:fourpointfour}
\end{align}
where $a_M$ is the first nonzero coefficient after $a_N$  \cite{16}.
Note that  since $[t] = L^2$, the maximum power of $F_{\mu\nu}$ in $a_M$ is $M + 2$ so  that the truncation error in (\ref{eq:fourpointtwo}) never exceeds $\mathrm O(e^2)$.

From (\ref{eq:fourpointthree}), (\ref{eq:fourpointfour}) and the result
\begin{align}
    \int_{1/e\mathscr F}^\infty \frac{\mathrm dt}{t}\,
    e^{-tm^2}=\ln\left(\frac{e\mathscr F}{m^2}\right)-\gamma+R,
    \tag{4.5}
    \label{eq:fourpointfive}
\end{align}
where  $\gamma = 0.5772\ldots$ is Euler's constant and $0<|R|<m^2/(e\mathscr F)$, obtain from  (\ref{eq:fourpointtwo}) for $e\gg1$
\begin{align}
    \ln\mbox{$\det_\mathrm{SQED}$}\ge
    -\frac{e^2\|F\|^2}{192\pi^2}\ln\left(\frac{e\mathscr F}{m^2}\right)
    +\mathrm O(e^2).
    \tag{4.6}
    \label{eq:fourpointsix}
\end{align}

We chose $e\mathscr F$ as the scaling parameter in (\ref{eq:fourpointtwo}).
Why not $e^\alpha\mathscr F$?
We set $\alpha= 1$ firstly because we remarked in Sec.II that $e$ always  multiplies $F_{\mu\nu}$   so that large amplitude variations of $F_{\mu\nu}$  can  just as well be studied in the strong coupling limit; setting $\alpha\neq1$ breaks this correspondence.
Secondly, if  $\alpha >1$ then the  lower bound in (\ref{eq:fourpointsix}) would be more negative, hence not optimal.
If $\alpha <1$ one gets a better bound in (\ref{eq:fourpointsix}) but the truncation  error in (\ref{eq:fourpointtwo}) increases faster than $e^2$ for terms of $\mathrm O(F^4)$ and  higher order.
So    $\alpha=1$ is the unique choice. The scaling parameter is further discussed in Sec.VI A.

The lower bound in (\ref{eq:fourpointsix}) is related to and in argeement with  the constant magnetic field growth of scalar QED's effective action  \cite{17} 
\begin{align}
    S_\mathrm{SQED}=-\ln\mbox{$\det_\mathrm{SQED}$}=\frac{B^2V}{96\pi^2}e^2\ln\left(\frac{eB}{m^2}\right)+\mathrm O(e^2).
    \tag{4.7}
    \label{eq:fourpointseven}
\end{align}
where $V$ is a four-dimensional volume cutoff.

This completes the discussion of the growth of the first term  in (\ref{eq:twopointsix}) and (\ref{eq:threepointnine}).
We now turn to the all-important second term.

\section{Strong Coupling Behavior of $\boldsymbol{\det_3}$}
\subsection{Paramagnetic property of $\boldsymbol{\det_3}$}
In Appendix A it is shown that $\Delta_A^{1/2}\sigma F\Delta_A^{1/2}\equiv T$ belongs to the  trace ideal $\mathscr I_3$ for $F_{\mu\nu}\in\cap_{p>2}  L^p (\mathbb R^4)$ and $m>0$.
This means that $T$ is a  compact operator that, in our case, maps $L^2(\mathbb R ^4)$ into itself.
Being compact its eigenvalues, $\{\lambda_n\}_{n=1}^\infty$, are discrete, and each  has finite multiplicity.
We order the $\lambda_n$    by $|\lambda_1|\ge|\lambda_2|\ge\ldots>0$.
 Because $T\in\mathscr I_3$     the eigenvalues  $\lambda_n\rightarrow 0$ and satisfy
 \begin{align}
     \sum_{n=1}^\infty|\lambda_n|^3<\infty.
     \tag{5.1}
     \label{eq:5.1}
 \end{align}
 Finally, $\ln\mbox{$\det_3$} (1+T)$ is gauge invariant (Appendix D) and satisfies  by (\ref{eq:threepointseven})
 \begin{align}
     \begin{split}
         &\ln\mbox{$\det_3$}\left(1+\Delta_A^{1/2}\frac{1}{2}e\sigma F\Delta_A^{1/2}\right)\\
         &\quad= \ln\det\left[ \left( 1+T \right)\exp\left( -T+\frac{1}{2} T^2\right) \right]\\
         &\quad= \mathrm{Tr}\left[ \ln(1+T)-T+\frac{1}{2}T^2 \right]\\
         &\quad= \sum_{n=1}^\infty\left[ \ln(1+\lambda_n)-\lambda_n+\frac{1}{2}\lambda_n^2 \right].
     \end{split}
     \tag{5.2}
     \label{eq:5.2}
 \end{align}
 In Appendix D it is shown that for every eigenstate of $T$ with  eigenvalue $\lambda_n$  there is another with eigenvalue $-\lambda_n$.
 Therefore,  (\ref{eq:5.2}) becomes
 \begin{align}
 \begin{split}
         &\ln\mbox{$\det_3$}\left(1+\Delta_A^{1/2}\frac{1}{2}e\sigma F\Delta_A^{1/2}\right)\\
         &\quad=\sum_{n=1}^\infty\left[ \ln(1-\lambda_n^2)+\lambda_n^2 \right],
  \end{split}
  \tag{5.3}
  \label{eq:5.3}
 \end{align}
 where the sum is over positive eigenvalues. We will see in Sec.VII B that the condition on $F_{\mu\nu}$ can be relaxed somewhat.

 Since $\ln\det_3$   is real and finite then  $\lambda_n<1$ for all $n$.
Hence,
\begin{align}
    \ln\mbox{$\det_3$}\left( 1+\Delta_A^{1/2}\frac{1}{2}e\sigma F\Delta_A^{1/2} \right)\le 0,
    \tag{5.4}
    \label{eq:5.4}
\end{align}
 since $\ln(1-x^2 )+x^2  \le0$ for $0\le x<1$.
This inequality has a physical origin.
Referring to (\ref{eq:threepointfive}) and (\ref{eq:threepointsix}) and simplifying exactly as outlined in  Appendix C for the function $\Pi$   we obtain 
\begin{align}
\begin{split}
    \ln\mbox{$\det_3$}&=\int_{0}^{\infty}\frac{\mathrm dt}{t}e^{-tm^2}\mathrm{Tr}\Biggl[ e^{-tD^2} 
     -e^{-t\left( D^2+\frac{1}{2}e\sigma F \right)}\\
    &+\frac{e^2}{8}te^{-tD^2/2}\sigma F\Delta_A^{1/2}\Delta_A^{1/2}\sigma Fe^{-tD^2/2}\Biggr].
    \end{split}
    \tag{5.5}
    \label{eq:5.5}
\end{align}
That $\ln\det_3  < 0$ is now seen as a consequence of the paramagnetism of  a charged spin-1/2 fermion in a static, four-dimensional  magnetic field $F_{\mu\nu}$: on average its energy levels are lowered by $F_{\mu\nu}$.
This  is made more precise by a version of the Peierls-Bogoliubov inequality derived from Klein's inequality \cite{18,19,20}:
\begin{align}
 \mathrm{Tr}\Big(e^{-t(P-eA)^2} - e^{-\left[ \left( P-eA \right)^2+\frac{1}{2}e\sigma F \right]t}\Big)\le 0 .
  \tag{5.6}
    \label{eq:5.six}
\end{align}
The last term in (\ref{eq:5.5}) has been purposely written in the form $U^\dagger U$ and is  therefore positive.
Nevertheless, it is dominated by the paramagnetism  of charged fermions through (\ref{eq:5.six}) which drives the integral in (\ref{eq:5.5}) to a negative  value.
This explains the second term in (\ref{eq:twopointsix}).

\subsection{Lower bound on $\boldsymbol{\ln\det_3}$ in the absence of zero modes}
The eigenvalues in (\ref{eq:5.3}) are obtained from
\begin{align}
    \frac{e}{2}\Delta_A^{1/2}\sigma F\Delta_A^{1/2}\varphi_n=-\lambda_n\varphi_n,
    \tag{5.7}
    \label{eq:5.seven}
\end{align}
where  $\varphi_n\in L^2$.
Let  $\Delta_A^{1/2}\varphi_n=\psi_n$     and obtain
\begin{align}
    \left[ (P-eA)^2+\frac{e}{2\lambda_n}\sigma F \right]\psi_n=-m^2\psi_n,
    \tag{5.8}
    \label{eq:5.8}
\end{align}
  where $\psi_n\in L^2$  as shown at the end of Appendix A.
  Eq. (\ref{eq:5.8})  illustrates the role of the eigenvalues $\{ \lambda_n \}_{n=1}^{\infty}$ as coupling constants whose discrete values result in bound states  with energy $-m^2$   for a fixed value of $e$.

  Because $\gamma_5$ commutes with $\sigma$, an eigenstate $\psi_n$ of (\ref{eq:5.8}) has  definite chirality.
In the representation (D7) $\gamma_5$   is diagonal  with elements $\pm\mathbbm 1_2$, and so we need only deal with the  two-dimensional  chirality eigenstates $\psi_n^\pm$.

      We note that each eigenvalue $\lambda_n(e)$ is a bounded function of $e$  as required by $|\lambda_n(e)|<1$  for all finite values of of $e$.
This is  illustrated by the constant field case:
\begin{align}
    |\lambda_n|=\frac{|eB|}{(2n+1)|eB|+m^2},\quad n=0,1,\ldots
    \tag{5.9}
    \label{eq:5.nine}
\end{align}
Therefore, the series in (\ref{eq:5.3}) will tend to an $e$-independent limit  for $e\gg1$ unless the degeneracy of the eigenvalues increases with $e$.
 The special case of a zero mode supporting background potential that  allows $|\lambda_n|$ to approach $1$ arbitrarily closely for $e\gg1$ will be  considered in the next section.

      To bound $\ln\det_3$   for $e\gg1$ we will first estimate the eigenvalue  degeneracy for the most symmetric case of an $\mathrm O(2)\times\mathrm O(3)$ background  field.
This estimate will place an upper bound on the eigenvalue  degeneracy of any random field.
The $\mathrm O(2)\times\mathrm O(3)$ symmetric fields  have the standard form \cite{21,22,23}
\begin{align}
    A_\mu(x)=M_{\mu\nu}x_\nu a(r),
    \tag{5.10}
    \label{eq:5.10}
\end{align}
where $M_{\mu\nu}$   is the antiself-dual antisymmetric matrix with  nonvanishing elements $M_{12}   = M_{30}   = 1$ and $r^2  = x_\mu^2$.
Alternatively $M$  may be replaced with the self-dual antisymmetric matrix $N$ with  nonvanishing elements $N_{03}   = N_{12}   = 1$.

Choosing the matrix $M$ the eigenstates of (\ref{eq:5.8}) have the form \cite{23}
\begin{widetext}
    \begin{align}
        \psi_n=r^{-2j-3/2}
        \left( \begin{array}{c}
            \mathscr D^j_{M-\frac{1}{2},m}(x)\rho_1(r)  \\
            \mathscr D^j_{M+\frac{1}{2},m}(x)\rho_2(r)  \\
            (j+m)^{\frac{1}{2}}r\rho_3(r)\mathscr D^{j-\frac{1}{2}}_{M,m-\frac{1}{2}}(x) 
            - (j-m+1)^{\frac{1}{2}} (\rho_4(r)/r) \mathscr D^{j+\frac{1}{2}}_{M,m-\frac{1}{2}}(x) \\
            (j-m)^{\frac{1}{2}}r\rho_3(r)\mathscr D^{j-\frac{1}{2}}_{M,m+\frac{1}{2}}(x)  
            + (j+m+1)^{\frac{1}{2}} (\rho_4(r)/r) \mathscr D^{j+\frac{1}{2}}_{M,m+\frac{1}{2}}(x) \\
        \end{array}\right),
        \tag{5.11}
        \label{eq:5.11}
    \end{align}
\end{widetext}
where $\mathscr D^j_{m_1m_2}(x)$ are the four-dimensional rotation matrices \cite{23,24,25}  normalized so that
\newpage
\begin{align}
    \int \mathrm d\Omega_4\,\mathscr D_{m_1m_2}^{j*}(x)\mathscr D_{m_3m_4}^{j'}(x)=\delta_{jj'}\delta_{m_1m_3}\delta_{m_2m_4}\frac{2\pi^2r^{4j}}{2j+1},
    \tag{5.12}
    \label{eq:5.12}
\end{align}
and where $2j = 0, 1,\ldots$; $-j \le m_i \le j$.
This paper follows the  conventions of \cite{23,24}; closely related ones appear in \cite{25}.
The  index $n$ has been omitted from $\rho_i$.
Inserting the two positive  chirality components of (\ref{eq:5.11}) into (\ref{eq:5.8}) results in the following  equations for $\rho_{1,2}$ \cite{24},
\begin{widetext}
    \begin{align}
        \left[ -\frac{\mathrm d^2}{\mathrm dr^2} + \frac{(2j+1)^2-\frac{1}{4}}{r^2}
        + (4M\mp2)ea + e^2r^2a^2 \pm \frac{e}{\lambda^+_n}(4a+r\frac{\mathrm da}{\mathrm dr})\right]\rho_{1,2}
        =-m^2\rho_{1,2},
        \tag{5.13}
        \label{eq:5.13}
    \end{align}
\end{widetext}
where the upper (lower) sign applies to $\rho_1$   ($\rho_2$), and $\lambda_n^+$  denotes  a positive chirality eigenvalue.
Since $(P-eA)^2 + \frac{e}{2}\sigma F \ge 0$ it is the     $\lambda_n^+$-dependent terms in (\ref{eq:5.13}) that are responsible for bound states  at $-m^2$.
There is a sequence of eigenvalues $1>\lambda_1^+ \ge \lambda_2^+  \ge \ldots>0$ dependent  on $e$, $j$, $M$, $m$, and the parameters specifying $A_\mu$  that result in  bound state solutions of (\ref{eq:5.13}).
They are independent of the quantum number $m$ in (\ref{eq:5.11}), resulting in a $(2j+1)$-fold degeneracy. Inspection of (\ref{eq:5.13}) indicates that in the positive chirality sector
\begin{align}
   \begin{split}
       \frac{1}{2}(\sigma F)^+&= \left(4a+r\frac{\mathrm da}{\mathrm dr}\right)\sigma_3\\
        &\equiv V(r)\sigma_3 .
   \end{split}
    \tag{5.14}
    \label{eq:5.14}
\end{align}
In general the degeneracy of the level at $-m^2$ has contributions from both $\rho_1$ and $\rho_2$. Consider $\rho_1$. Assume that $a$ and $a'$ are bounded functions of $r$. Inclusion of zero modes requires 
  $\lim_{r\rightarrow\infty}r^2 a = \nu$, where we may assume $\nu>0$ as discussed in Sec.C below. Then $r^2 V(r)$ is a bounded function of r and
 \begin{align}
    \inf\left[ r^2V(r) \right]=-K_1>-\infty.
     \tag{5.15}
    \label{eq:5.15}
\end{align}
The $\lambda_n^+$-independent terms on the left-hand side of (\ref{eq:5.13}) form a positive operator whose controlling parameter is $j$ for fixed $e$.
Thus a bound state at $-m^2$ can exist only if
\begin{align}
   \left( 2j+1 \right)^2<\frac{e}{\lambda_n^+}K_1+\frac{1}{4}.
   \tag{5.16}
   \label{eq:5.16}
\end{align}
This is a necessary condition but obviously not a sufficient one.
The maximum allowed value of $j$ for all finite values of $m^2$ and a fixed value of $M$ is  $J_1  < \left( \frac{eK_1}{4\lambda_n^+}+\frac{1}{16} \right)^\frac{1}{2}-\frac{1}{2}$.
Hence, the maximum degeneracy $\mu_{1n}^+$ of  eigenvalue $\lambda_n^+$ associated with $\rho_1$ for $\frac{eK_1}{\lambda_n^+} \ge 1$ is
\begin{align}
   \mu_{1n}^+=\sum_{j=0,\frac{1}{2},\dots}^{J_1}(2j+1) < 2 \left[  \left(\frac{eK_1}{4 \lambda_n^+}\right) ^{\frac{1}{2}} +1 \right]^2 .
    \tag{5.17}
    \label{eq:5.17}
\end{align}
For the other positive chirality state $\mathscr D_{M+\frac{1}{2},m}^{j}\rho_2/r^{2j+\frac{3}{2}}$ inspection  of (\ref{eq:5.13}) indicates that the bound state at $-m^2$  acquires an additional maximal degeneracy  $\mu_{2n}^+$ satisfying the bound in (\ref{eq:5.17}) with $K_1$  replaced with $K_2=\sup(r^2V(r))<\infty$.
It may happen that either $\rho_1$ or $\rho_2 $ has no bound states at $-m^2$.

Is the dependence of $\mu_{1n}^+$, $\mu_{2n}^+$ on  $\lambda_n^+$ reasonable?
As $\lambda_n^+\searrow0$  the potential wells in $\pm\frac{e}{\lambda_n^+}V(r)$ deepen, increasing the probability  that such wells can support a bound state at $-m^2$.
As the wells  deepen the centrifugal barrier term in (\ref{eq:5.13}) can increase, thereby  allowing larger values of $j$ and hence higher degeneracy, consistent  with our result (\ref{eq:5.17}).

      In the negative chirality sector
\begin{align}
    \frac{1}{2}(\sigma F)^-=\left( \begin{array}{cc}
        -\mathscr D_{00}^1  &   \sqrt 2\mathscr D_{01}^{1*} \\
        \sqrt 2\mathscr D_{01}^1    &   \mathscr D_{00}^1
    \end{array}\right)
    \frac{1}{r}\frac{\mathrm da}{\mathrm dr},
    \tag{5.18}
    \label{eq:5.8een}
\end{align}
where $\mathscr D_{00}^1=x_0^2+x_3^2-x_1^2-x_2^2$  and $\mathscr D_{01}^1=-\sqrt 2(x_0+ix_3)(x_2-ix_1)$.
Insertion of (\ref{eq:5.8een}) and the two negative chirality components  of (\ref{eq:5.11}) in (\ref{eq:5.8}) results in coupled equations for $\rho_3$ and $\rho_4$:
\begin{widetext}
    \begin{align}
        \left( -\frac{\mathrm d^2}{\mathrm dr^2} + \frac{4j^2-\frac{1}{4}}{r^2}
        + 4Mea + e^2r^2a^2\right)\rho_3
        + \frac{e}{\lambda_n^-}ra'\left( 
        \sqrt{1-\frac{M^2}{(j+\frac{1}{2})^2}}\rho_4
        + \frac{M}{j+\frac{1}{2}}\rho_3\right)
        = -m^2\rho_3
        \tag{5.19}
        \label{eq:5.19}\\
        \left( -\frac{\mathrm d^2}{\mathrm dr^2} + \frac{4(j+1)^2-\frac{1}{4}}{r^2}
        + 4Mea + e^2r^2a^2\right)\rho_4
        + \frac{e}{\lambda_n^-}ra'\left( 
        \sqrt{1-\frac{M^2}{(j+\frac{1}{2})^2}}\rho_3
        - \frac{M}{j+\frac{1}{2}}\rho_4\right)
        = -m^2\rho_4 .
        \tag{5.20}
        \label{eq:5.20}
    \end{align}
\end{widetext}
These equations can be decoupled for large j by a unitary transformation $U$ on $\rho_3$, $\rho_4$. Let $U\rho=\varphi$ with $U_{33}=U_{44}=(\frac{1+M}{(j+\frac{1}{2})})^{\frac{1}{2}}/\sqrt2$
and $U_{34}=-U_{43}=(\frac{1-M}{(j+\frac{1}{2})})^{\frac{1}{2}}/\sqrt2$ so that the coupled terms in (\ref{eq:5.19}),(\ref{eq:5.20})proportional to $e/\lambda_n^-$ are transformed to $(e/\lambda_n^-)ra'\sigma_3\varphi$.
Comparing this with (\ref{eq:5.13}) the same analysis used in the positive chirality case applies here. Thus, following (\ref{eq:5.17}) the maximum degeneracies
$\mu_{3n}^-$, $\mu_{4n}^-$associated with the bound states $\varphi_3$, $\varphi_4$ at $-m^2$ are bounded by $eK/\lambda_n^-$, where $K$ is an $e$-independent constant.
This assumes $e/\lambda_n^->>1$ corresponding to large j.

We emphasize that the estimated maximum degeneracies above are for one level at $-m^2$. They are not an estimate of the number of bound states at energy $\le -m^2$ which is expected to vary as $e^2$ for $F_{\mu\nu}\in L^2$ by theorem 2.15 in \cite{26}.

We now have estimates for the maximum degeneracy of eigenvalues  $\lambda_n^{\pm}$  obtained  from (\ref{eq:5.8}) for the most symmetric  admissible background field given by (\ref{eq:5.10}).
The above results  place an upper bound on the eigenvalue degeneracy $\mu_n$ of any  admissible random field, namely for $e\gg1$
\begin{align}
    \mu_n(e)<\frac{ec}{\lambda_n} ,
    \tag{5.21}
    \label{eq:5.21}
\end{align}
where $\lambda_n$ is one of the random field's eigenvalues obtained from  (\ref{eq:5.8}), and $c$ is $e$-independent.
The $1/\lambda_n$   dependence of its right-hand side is important  because it results in the convergent series $\sum_{n>N}^\infty \lambda_n^3$ in (\ref{eq:5.23})  below, whatever the field may be.

Consider the series in (\ref{eq:5.3}) and divide it into $\sum_{n=1}^N+\sum_{n>N}^\infty$,  where $\lambda_n^2<\frac{1}{2}$ for $N>n$, $N$ sufficiently large.
Note in this case that
\begin{align}
    \frac{1}{2}\le\left|\frac{\ln(1-\lambda_n^2)+\lambda_n^2}{\lambda_n^4}\right|
    < 1.
    \tag{5.22}
    \label{eq:5.22}
\end{align}
Thus for any admissible random field, excluding those that support  a zero mode, there follows from (\ref{eq:5.3}), (\ref{eq:5.21}), and (\ref{eq:5.22})
\begin{align}
\begin{split}
    &\left| \ln\mbox{$\det_3$}\left( 1+\Delta_A^\frac{1}{2}\frac{1}{2}e\sigma F\Delta_A^\frac{1}{2} \right)\right |\\
    &<\sum _{n=1}^N \left|\ln(1-\lambda_n^2)+\lambda_n^2\right| + \sum_{n>N}^{\infty}\lambda_n^4\\
    &<\sum_{n=1}^N\left|\ln(1-\lambda_n^2)+ \lambda_n^2\right|+ec \underset{\text{no degeneracy}}{\sum_{n>N}^{\infty}}\lambda_n^3 ,
    \end{split}
    \tag{5.23}
    \label{eq:5.23}
\end{align}
where the third line of (\ref{eq:5.23}) is valid when $e\gg1$.
In the absence of zero modes $\lim_{e\rightarrow\infty} \lambda_1< 1$ unlike the zero mode case discussed in Sec.C below.
By (\ref{eq:5.1})  the infinite series on the right converges.
Moreover, the $e\rightarrow\infty$  limit of this series is finite.
Thus, there is a number $M$ such  that for $n>M$, $\lambda_n  (e) < C_n (e)/n^{1/3+\epsilon}$,  $\epsilon>0$ and $C_n$  is a bounded function  of $n$ and $e$ with $\lim_{e\rightarrow\infty} C_n (e)< \infty$.
Otherwise    $\lambda_n<1$ for any $n$ cannot be  satisfied.
Accordingly, the right-hand series in (\ref{eq:5.23}) is  uniformly convergent in $e$ by the Weierstrass M test, allowing its  $e\rightarrow\infty$     limit to be taken term-by-term and establishing our claim.
The remaining series,  $\sum_{n=1}^N|\ln(1 - \lambda_n^2 ) + \lambda_n^2|$,  is obviously bounded by $e$ following (\ref{eq:5.21}), excluding zero modes.
Combining (\ref{eq:5.3}), (\ref{eq:5.21}), (\ref{eq:5.22}) and (\ref{eq:5.23}) gives in the absence of zero modes
\begin{align}
    0\ge\lim_{e\rightarrow\infty}\ln\mbox{$\det_3$}
    \left( 1+\Delta_A^\frac{1}{2}\frac{1}{2}e\sigma F\Delta_A^\frac{1}{2} \right)/e > -C,
    \tag{5.24}
    \label{eq:5.24}
\end{align}
where $C>0$ is an $e$-independent constant depending on the  specific background field. $C$ must be a linear function of $F_{\mu\nu}$ to preserve the correlation $eF_{\mu\nu}$.

\subsection{Zero modes}
Consideration is now given to potentials supporting $L^2$  zero  modes of the Dirac operator $\slashed P - e\slashed A$.
It is these potentials that  provide the mechanism governing the stability of QED and its  non-perturbative quantization.

      The relevance of zero modes to stability arises as follows.
      Suppose $A_\mu$   supports a zero mode, $\psi_{\mathrm{zero},n}$, where $n$ denotes the  quantum numbers required to specify it.
It is an $L^2 $ solution of
\begin{align}
    \left[ \left( P-eA \right)^2 + \frac{e}{2}\sigma F \right]\psi_{\mathrm{zero},n}=0,
    \tag{5.25}
    \label{eq:5.25}
\end{align}
obtained from (\ref{eq:5.8}) by setting  $\lambda_n= 1$, $m=0$.
We continue to assume  $\lambda_n> 0$ as discussed in Sec. V A.
Then (\ref{eq:5.25}) requires  $ \bra{\mathrm{zero},n}\sigma F\ket{\mathrm{zero},n}< 0$.
Refer to (\ref{eq:5.8}) and replace $\lambda_n$  with a general eigenvalue $\lambda$   and denote the corresponding  eigenstate by $\psi_{\lambda,n}$.
Assume $\bra{\lambda,n}\sigma F\ket{\lambda,n}<0$.
Then from (\ref{eq:5.8})  and (\ref{eq:5.25}) there follows
\begin{align}
    \frac{e}{2}\left( \frac{1}{\lambda}-1 \right)\bra{\mathrm{zero},n}\sigma F\ket{\lambda,n}=-m^2\braket{\textrm{zero},n}{\lambda,n}.
    \tag{5.26}
    \label{eq:5.26}
\end{align}
There is no \emph{a priori} reason why the two sides of (\ref{eq:5.26}) should  vanish if the quantum numbers of the two states are the same.
Based  on our limited knowledge of four-dimensional Abelian zero modes \cite{24}  they have a distinctive structure, and so the nonvanishing of  $\braket{\textrm{zero},n}{\lambda,n}$ distinguishes the eigenstate $\psi_{\lambda,n}$ --and its  eigenvalue $\lambda$-- from other eigenstates obtained from (\ref{eq:5.8}).

Divide (\ref{eq:5.26}) by $e$. For $e\gg1$ conclude that $\lambda$ has the form
\begin{align}
    \lambda=1-\delta(e,n,m,L,\dots) ,
    \tag{5.27}
    \label{eq:5.27}
\end{align}
where $0<\delta<1$  and that for fixed $m$, $\delta\searrow  0$ for $e\rightarrow \infty$.
$L$ is a parameter with the dimension of length introduced by $A_\mu$ that can combine  with $m$ to form a dimensionless $\delta$.
This result requires that the states $\psi_{\lambda,n}$ be in one-to-one correspondence with the zero modes $\psi_{\text{zero},n}$.
The eigenvalue $\lambda$ will be discussed  for an analytically solvable case in Sec. 5 E.

Insertion of (\ref{eq:5.27}) in (\ref{eq:5.3}) gives
\begin{align}
\begin{split}
&\ln\mbox{$\det_3$}=\sum_n\sigma_n\\
&\times\left[ -\ln\left(\frac{1-\delta}{\delta}\right)+\ln\left[(1-\delta)(2-\delta)\right]+(1-\delta^2)\right]+\dots,
\end{split}    
\tag{5.28}
    \label{eq:5.28}
\end{align}
where the remainder in (\ref{eq:5.28}) is the contribution from eigenvalues  bounded away from $1$ discussed in the previous section; $\sigma_n$ is the degeneracy of state $n$..
The sum in  (\ref{eq:5.28}) is over the quantum numbers specifying the zero modes of $A_\mu$.
Write (\ref{eq:5.26}) in the form
\begin{align}
\frac{1-\delta}{\delta}= \frac{e}{2m^2}\left|\frac{\langle\text{zero},n|\sigma F|\lambda,n\rangle}{\langle\text{zero},n|\lambda,n\rangle}\right|,
\tag{5.29}
    \label{eq:5.29}
\end{align}
where
\begin{align}
\left|\frac{\langle\text{zero},n|\sigma F|\lambda,n\rangle}{\langle\text{zero},n|\lambda,n\rangle}\right|\le K \mathscr F.
\tag{5.30}
    \label{eq:5.30}
\end{align}
Eq. (\ref{eq:5.30}) assumes $F_{\mu\nu}(x)$ is a bounded function in which case $K$ is an $e$-independent constant; $\mathscr F$ is the amplitude of $F_{\mu\nu}$ corresponding to the scaling parameter introduced in Sec.IV. Inserting (\ref{eq:5.29}) in (\ref{eq:5.28}) gives for $ e \rightarrow \infty$
\begin{align}
\begin{split}
&\ln\mbox{$\det_3$}= -\sum_n \sigma_n \\
&\times\left[\ln\left(\frac{e\mathscr F}{m^2}\right)+\ln\left|\frac{\langle\text{zero},n|\sigma F|\lambda,n\rangle/\mathscr F}{\langle\text{zero},n|\lambda,n\rangle}\right|-2\ln 2-1\right]\\
&+ \text{O(e)}.
    \end{split}
    \tag{5.31}
    \label{eq:5.31}
\end{align}
The $O(e)$ term is the contribution from the eigenvalues bounded away from 1 discussed in the previous section.
Since 
 \begin{align}
     \sum_n\sigma_n=\#\mbox{ zero modes supported by } A_\mu,
     \tag{5.32}
     \label{eq:5.32}
 \end{align}
 if the number of zero modes increases as $e^2$   or faster then the result (\ref{eq:5.31}) will  override the bound in (\ref{eq:5.24}) and possibly drive $\ln\det_\textrm{ren}$  in (\ref{eq:threepointnine})  negative.
Clearly, these considerations are highly relevant to  QED's non-perturbative quantization.

\subsection{Counting zero modes}
Following (\ref{eq:5.31}) and (\ref{eq:5.32}) it is of exceptional interest to  know the maximum number of zero modes a potential can support.
To  begin we focus on the most symmetric admissible potentials (\ref{eq:5.10}).
It is assumed that zero mode potentials within the class (\ref{eq:5.10})  will produce the maximum number due to their high symmetry  and hence large number of degenerate states $\psi_{\mathrm{zero},n}$.
As pointed out  in the previous section, eigenstates $\psi_{\lambda,n}$ of (\ref{eq:5.8}) with eigenvalue $\lambda$ given by (\ref{eq:5.27}) will be in one-to-one correspondence with the  states $\psi_{\mathrm{zero},n}$.
We would then expect that zero mode supporting  potentials with lesser symmetry will have their zero mode number  bounded by this most symmetric result.
It turns out that this  reasoning is not completely correct and that potentials with  lesser symmetry can compete with those in (\ref{eq:5.10}).
This is a huge  advantage for QED's stability.
We will begin with the potentials  (\ref{eq:5.10}) and then explain why this reasoning has to be modified.

The zero modes supported by the potentials in (\ref{eq:5.10}) have  been discussed in \cite{24}.
We continue to assume that $a$ and $a'$ are  bounded functions of $r$ and in addition $\lim_{r\rightarrow\infty} r^2 a = \nu$,  $\nu\neq 0$.
That is, $A_\mu$ must have a $1/r$ falloff.
This ensures that the global  chiral anomaly  $\mathcal A$ is nonvanishing:
\begin{align}
    \mathcal A = -\frac{1}{16\pi^2}\int\mathrm d^4x \,^\star F_{\mu\nu}F_{\mu\nu}=\pm\frac{\nu^2}{2},
    \tag{5.33}
    \label{eq:5.33}
\end{align}
where  $^\star FF =  \partial_\alpha (\epsilon_{\alpha\beta\mu\nu}     A_\beta F_{\mu\nu} )$.
The $+$($-$) sign in (\ref{eq:5.33}) results in the  case of matrix $M$ ($N$) defined under (\ref{eq:5.10}).
Without loss of  generality we will assume   $\nu> 0$.
The nonvanishing of $\mathcal A$ indicates  that $F_{\mu\nu}$ is not square-integrable. We repeat here that it is sufficient to assume $F_{\mu\nu}\in \cap_{p>2}L^p$ to define $\det_3$, and therefore it can accommodate zero modes.

Choosing the matrix $M$ in (\ref{eq:5.10}) it is found that only the  positive chirality sector has normalizable zero modes \cite{24}.
This  is a particular example of a vanishing theorem: all normalizable  zero modes of $\slashed D^2$  have only one chirality.
There is no such general  theorem in $\text{QED}_4$,  unlike the non-Abelian case \cite{27,28} and $\text{QED}_2$  \cite{29}.
 Up to a normalization constant these are \cite{24}
 \begin{align}
     \psi_\mathrm{zero}(x)=\mathscr D_{-j,m}^j(x)
     e^{-e\int_{r_0}^{r}\mathrm dr\, ra(r)}
     \left(\begin{array}{c}
        0\\
        1\\
        0\\
        0
     \end{array}\right)
     \tag{5.34}
     \label{eq:5.34}.
 \end{align}
 Here $\exp\left[-e \int_{r_0}^{r}\mathrm dr\,ra(r)\right] =  \rho_2$  in (\ref{eq:5.11}) when $M = -j-1/2$   and in (\ref{eq:5.13})  when in addition $m^2  = 0$ and    $\lambda_n^+= 1$.
 Eq. (\ref{eq:5.34}) and the assumption  $a(r) \underset{r\rightarrow\infty}{\sim}{\nu}/r^2$  indicate that $\psi^+$ is square-integrable provided  $e\nu  > 2j+2$.
 Following (\ref{eq:5.32}),
 \begin{align}
     \mbox{\# zero modes}=\sum_{j=0,\frac{1}{2},\dots}^{j_\mathrm{max}}(2j+1)
     =\frac{1}{2}[e\nu]^2-\frac{1}{2}[e\nu],
     \tag{5.35}
     \label{eq:5.35}
 \end{align}
 where $[x]$ is the greatest integer less than $x$.
 Using (\ref{eq:5.33}) for  $e\nu \gg1$,
 \begin{align}
     \mbox{\# zero modes}&=\frac{1}{2}(e\nu)^2+\mathrm O(e\nu)\\
     &=\frac{e^2}{16\pi^2}\left| \int\mathrm d^4x\, ^\star F_{\mu\nu}F_{\mu\nu}\right|
     + \mathrm O(e\nu).
     \tag{5.36}
     \label{eq:5.36}
 \end{align}
 If the matrix $M$ is replaced with $N$ in (\ref{eq:5.10}) the zero modes shift  to the negative chirality sector.
 Therefore, (\ref{eq:5.36}) includes this  case.

 Given another potential with lesser symmetry than $\mathrm O(2)\times\mathrm O(3)$  and having the same chiral anomaly we tentatively conclude that  its zero mode number is bounded by the right-hand side of (\ref{eq:5.36}).
 This assumes that all of the potential's zero modes have one  chirality only.

      More information about the zero mode number of less symmetric  potentials can be obtained from the index theorem for non-compact  Euclidean space-time \cite{30},
      \begin{align}
          n_+-n_--\frac{1}{\pi}\sum_{l}\left[ \delta_l^+(0) - \delta_l^-(0) \right]
          = - \frac{e^2}{16\pi^2}\int\mathrm d^4x\, ^\star F_{\mu\nu}F_{\mu\nu},
          \tag{5.37}
          \label{eq:5.37}
      \end{align}
  where $n_\pm$  is the number of positive/negative chirality $L^2$  zero modes; $\delta^\pm_l(0) \in(0, \pi ]$ are the zero energy scattering phase shifts for the  Hamiltonians $H_\pm  = \frac{1}{2}( 1 \pm \gamma_5)\slashed D^2$, and  $l$  denotes the quantum numbers  required to specify the phase shifts.
The sum over phase shifts  gives the fractional discrepancy between the index and the chiral  anomaly.
Consequently the sum in (\ref{eq:5.37}) grows more slowly than $e^2$  for $e\gg1$.
Based on (\ref{eq:5.37}) if there were a general vanishing theorem  for $\text{QED}_4$ then the $\mathrm O(2)\times\mathrm O(3)$ result in (\ref{eq:5.36}) would continue to  hold for potentials with lesser symmetry.
This perhaps  counterintuitive conclusion that two potentials with the same  chiral anomaly--one with maximal symmetry, the other with lesser  symmetry-- have the same number of zero modes is related  to their common asymptotic behavior.
Without a vanishing theorem  (\ref{eq:5.37}) implies that the total number of zero modes may exceed the  chiral anomaly.
Summarizing,
\begin{align}
    \begin{split}
        &\mbox{\# zero modes supported by }A_\mu\\&\quad\ge
        \frac{e^2}{16\pi^2}\left| \int\mathrm d^4x\, ^\star F_{\mu\nu}F_{\mu\nu}\right|+\Delta,
    \end{split}
   \tag{5.38}
    \label{eq:5.38}
\end{align}
 with the inequality applying in the absence of a vanishing theorem  and   $\Delta/e^2\rightarrow  O$ for $e\rightarrow\infty$.

Insertion of (\ref{eq:5.38}) in (\ref{eq:5.31}) gives with (\ref{eq:5.32}) 
 \begin{align}
     \begin{split}
         &\ln\mbox{$\det_3$}\\
         &\quad\le -\frac{1}{16\pi^2}\left|
         \int\mathrm d^4x\,^\star F_{\mu\nu}F_{\mu\nu}\right| e^2\ln \left(\frac{e \mathscr F}{m^2}\right) + R,
     \end{split}
     \tag{5.39}
     \label{eq:5.39}
 \end{align}
 with $R/(e^2 \ln e)\rightarrow   0$ for $e\rightarrow\infty$, in which case the bound in (\ref{eq:5.24})  is overridden.
 As noted in Sec.A the negative sign in (\ref{eq:5.39}) is  a consequence of the paramagnetism of a charged spin-$\frac{1}{2}$   fermion in a  static, four-dimensional magnetic field.

\subsection{Eigenvalue $\lambda$}
Because of the possible far-reaching implications of (\ref{eq:5.39})  for the non-perturbative quantization of QED and the electroweak  model it is important to have an analytic calculation of the  eigenvalue  $\lambda$  in (\ref{eq:5.27}) for a few special cases to show that the formalism outlined in Secs. C and D can be  implemented.

      We consider a class of maximally symmetric zero mode supporting  potentials (\ref{eq:5.10}) with profile function
      \begin{align}
          a(r)=\begin{cases}
              \frac{C}{R^2}\left( \frac{r}{R} \right)^{\epsilon-2} +
              \frac{(2-\epsilon)C-2\nu}{R^3}r + \frac{(\epsilon-3)C + 3\nu}{R^2},&r\le R\\
              \frac{\nu}{r^2},&r>R
          \end{cases}
          \tag{5.40}
          \label{eq:5.40}
      \end{align}

      It is constructed so that $a$ and $a'$, and hence $F_{\mu\nu}$   are continuous  at $r = R$.
      The parameter   $\epsilon\ge 2$ to ensure that $F \in\cap_{p>2}  L^p$.
 The constant $C$ can be positive or negative, and we continue to  assume   $\nu> 0$.

 As noted in Sec. D the $L^2$  zero modes of (\ref{eq:5.25}) reside in the  positive chirality sector with $M = -j-\frac{1}{2}$   for the potentials (\ref{eq:5.10}).
 A $L^2$  solution of (\ref{eq:5.8}) originating from the zero mode (\ref{eq:5.34}) is
 \begin{align}
     \psi_\lambda(x)=\mathscr D_{-jm}^j(x)\frac{f(r)}{r^{2j+\frac{3}{2}}}
     \left(\begin{array}{c}
         0 \\
         1 \\
         0 \\
         0 
     \end{array}\right),
     \tag{5.41}
     \label{eq:5.41}
 \end{align}
where $f\equiv\rho_2$     in (\ref{eq:5.13}) now satisfies
\begin{widetext}
    \begin{align}
        \left[ \frac{\mathrm d^2}{\mathrm dr^2}
            + \frac{\frac{1}{4}-(2j+1)^2}{r^2}
            + 4\left(j+\frac{1}{\lambda}\right)ea
            -e^2r^2a^2
        + \frac{e}{\lambda}r\frac{\mathrm da}{\mathrm dr}\right]f
        =m^2f,
        \tag{5.42}
        \label{eq:5.42}
    \end{align}
with eigenvalue $\lambda$ given (\ref{eq:5.27}) when $e\gg1$.
For $r>R$ let $f = r^{\frac{1}{2}} g$ so  that (\ref{eq:5.42}) becomes
    \begin{align}
        g''+\frac{1}{r}g'
        - \left( m^2 + \frac{(2j+1-e\nu)^2+2\left(1-\frac{1}{\lambda}\right)e\nu}{r^2} \right)g
        = 0,
        \tag{5.43}
        \label{eq:5.43}
    \end{align}
\end{widetext}
whose decaying solution is the modified Bessel function $K_\alpha  (mr)$  with
\begin{align}
    \alpha=\left[ (2j+1-e\nu)^2+2\left( 1 - \frac{1}{\lambda} \right)e\nu \right]^{\frac{1}{2}}.
    \tag{5.44}
    \label{eq:5.44}
\end{align}
The eigenvalue $\lambda$ is fixed by the boundary condition at $r=R$:
\begin{align}
    \frac{Rf'(R)}{f(R)}=\frac{1}{2}+\frac{RK_\alpha'(mR)}{K_\alpha(mR)}.
    \tag{5.45}
    \label{eq:5.45}
\end{align}
The left-hand side of (\ref{eq:5.45}) is calculated from the solution of  (\ref{eq:5.42}) for $0\le r\le R$.

      The analysis simplifies by assuming $mR\ll1$.
      Let $e\nu  = N + \Delta$,  $N = 2, 3,\dots$; $0< \Delta<1$, $j = 0,\frac{1}{2},\dots,j_\mathrm{max}$    with $j_\mathrm{max}   = (N-2)/2$ since $L^2$  zero modes exist only for $e\nu  > 2j + 2$.
      It is known that $\det_\mathrm{ren}$    has  a branch point in $m$ beginning at $m=0$ \cite{24} which is evident by the  presence of $K_\alpha$   in (\ref{eq:5.45}).
      This leads to the following small mass  expansions for $j = 0,\frac{1}{2},\dots, j_\mathrm{max}-\frac{1}{2}$   and     $\alpha_0=e\nu   -2j - 1 > 2$,
      \begin{align}
          f&=Bf_0\left(1+m^2f_2+m^4f_4+\mathrm O\left(m^{2\alpha_0}\mbox{ or }m^6\right)\right),
          \tag{5.46}
          \label{eq:5.46}\\
          \lambda &=1-m^2\delta_2 - m^4\delta_4 + \mathrm O\left(m^{2\alpha_0}\mbox{ or }m^6\right);
          \tag{5.47}
          \label{eq:5.47}
      \end{align}
      for $j = j_\mathrm{max}$, $1 <  \alpha_0 < 2$,
      \begin{align}
          f&=Bf_0\left( 1+m^2f_2+m^{2\alpha_0}f_{2\alpha_0}+\mathrm O \left( m^4 \right) \right)
          \tag{5.48}
          \label{eq:5.48}\\
          \lambda&=1-m^2\delta_2 - m^{2\alpha_0}\delta_{2\alpha_0} + \mathrm O\left(m^4\right);
          \tag{5.49}
          \label{eq:5.49}
      \end{align}
      where $\alpha_0$   is the $m = 0$ term in the expansion of $\alpha$ in (\ref{eq:5.44}), and $B$  is a normalization constant.
The expansion of $\delta$ in (\ref{eq:5.27}), (\ref{eq:5.47})and(\ref{eq:5.49}) in  powers of $m$ must begin at $m^2$  to be consistent with the boundary  condition (\ref{eq:5.45}).
For all cases there is a $\mathrm O(m^2)$ term in the  expansions of $f$ and  $\lambda$.
The case $e\nu   = 3, 4,\dots$ is commented on  in Appendix E.
Here $f_0$  is the solution of (\ref{eq:5.42}) when $m = 0$,    $\lambda= 1$ and $0\le r \le R$,
\begin{align}
    f_0=r^{2j+\frac{3}{2}}e^{-e\int_{0}^{r}\mathrm ds\,sa(s)} ,
    \tag{5.50}
    \label{eq:5.50}
\end{align}
With these expansions the two sides of (\ref{eq:5.45}) can be matched in  powers of $m$ to obtain $\lambda$.
The calculation is outlined in  Appendix E.

      For $mR<<1$, $e\nu  > 2j + 2$ and $e\gg1$ the calculation in Appendix E  gives, following (E11) and (E12),
      \begin{align}
          \lambda=1-\frac{2m^2/e}{\|(\sigma F(r_0))^+\|_1}
          (1+\mathrm O(1/e))
          + \mathrm O\left(\frac{m^4R^4}{e^2}\right),
          \tag{5.51}
          \label{eq:5.51}
      \end{align}
      where $(\sigma F)^+$  is the positive chirality component of   $\sigma F$ in (\ref{eq:5.14})  that is responsible for the existence of zero modes, and $r_0$  is the  unique root in the interval $0<r <R$ of 
      \begin{align}
          4j+3-2er^2a(r)=0.
          \tag{5.52}
          \label{eq:5.52}
      \end{align}
      Here  $\|(\sigma F)^+\|_1$ is the spin trace norm of $(\sigma F)^+$  defined for an  operator $A$ by $\|A\|_1=\mathrm{Tr}(A^\dagger A)^{1/2}$.
Because $(\sigma  F)^+$  obtained from (\ref{eq:5.14})  and (\ref{eq:5.40}) is a smooth function,   $\lambda$ is a slowly varying function of $j$  since $\mathrm dr_0 /\mathrm dj = \mathrm O(1/e)$ from (\ref{eq:5.52}).
For this special case we can count zero modes following (\ref{eq:5.35}), (\ref{eq:5.36}) and rewrite (\ref{eq:5.39}) as an  equality. To leading order in $m^2/e$, $\delta$ in (\ref{eq:5.27}) can be read off from (\ref{eq:5.51}). This fixes the argument of the logarithm in (\ref{eq:5.28}) precisely:

\begin{align}
\begin{split}
&\ln\mbox{$\det_3$}\\
&=-\sum_{j=0}^{j_{\text{max}}}(2j+1)\left[\ln\left(\frac{e \|(\sigma F(r_0(j)))^+\|_1}{2m^2}\right)+\text{O(1)}\right]+R_1 ,
\end{split}
\tag{5.53}
        \label{eq:5.53}
\end{align}
where $j_{\text{max}}=[e\nu]/2-1$ and $\text{lim}_{e \rightarrow \infty}R_1/(e^2\ln e)=0$. The remainder $R_1$ includes contributions to $\mbox{$\det_3$}$ from eigenvalues bounded away from 1 as discussed in Sec.B. Defining an average $F_{\mu\nu}, \mathscr F$, by

\begin{align}
\sum_{j=0}^{j_{\text{max}}}(2j+1) \ln \|(\sigma F(r_0(j)))^+\|_1 \left/ {\sum}_{j=0}^{j_{\text{max}}}(2j+1)\right. \equiv \ln \mathscr F
      \tag{5.54}
        \label{eq:5.54}
\end{align}
obtain from (\ref{eq:5.35}) and (\ref{eq:5.36}) for $e\gg1$
\begin{align}
\begin{split}
        &\ln\mbox{$\det_3$}\\
        & =-\frac{e^2}{16\pi^2}\left|\int\mathrm d^4x\,^\star F_{\mu\nu}F_{\mu\nu}\right|
        \left[\ln\left(\frac{e \mathscr F}{2 m^2}\right)+\text{O(1)}\right]+R_2,
         \end{split}     
        \tag{5.55}
        \label{eq:5.55}
         \end{align}
where $R_2$ contains a $\text{O}(\text{e}\nu\ln(e\mathscr F))$ term from the $\text{O}(\text{e}\nu)$ residue in (\ref{eq:5.36}) and satisfies the same limit as $R_1$. The result (\ref{eq:5.55}) overrides the bound (\ref{eq:5.24}).

\def\dt{\text{d}t}
\def\dfour{\text{d}^4}
\def\TR{\mathinner{\mathrm{Tr}}}

\section{Charge renormalization term in $\boldsymbol{\ln\det_\mathrm{ren}}$}
\subsection{ Scaling parameter}
Consider the last contribution to $\ln\det_\mathrm{{ren}}$ in \eqref{eq:twopointsix} and \eqref{eq:threepointnine}, here designated as
\begin{align}
    \Pi = e^2 \int_0^{\infty} \dt e^{-tm^2} \left[ \frac{\norm{F}^2}{32\pi^2 t} - \frac{1}{2} \TR \left( e^{-tD^2} F_{\mu \nu} \Delta_A F_{\mu \nu} \right) \right].
   \tag{6.1}
   \label{eq:6.1}
\end{align}
It is not obvious what to call the right-hand side of \eqref{eq:6.1}, but since $e^2 \norm{F}^2  /(32 \pi^2  t)$ is part of the on-shell charge renormalization
subtraction in $\ln\det_\mathrm{{ren}}$ it will be referred to as the charge renormalization term. As in Sec.IV break the integral in \eqref{eq:6.1} into $\int_{0}^{1/e\mathscr{F}}$ and  $\int_{1/e\mathscr{F}}^{\infty}$, where $\mathscr{F}$ fixes the scale of the amplitude of $F_{\mu \nu}$.
Then $\Pi = I_1 + I_2 + I_3$, where 
\begin{align}
    I_1 &= \frac{e^2 \norm{F}^2}{32\pi^2} \int_{1/e \mathscr{F}}^{\infty} \frac{\dt}{t}  e^{-tm^2},
   \tag{6.2}
   \label{eq:6.2}\\ 
    \nonumber
    I_2 &= \frac{e^2}{32} \int_0^{1/e \mathscr{F}}\dt e^{-tm^2} \\  
     &\times \left[ \frac{\norm{F}^2}{\pi^2 t}-16 \TR \left( e^{-tD^2} F_{\mu \nu} \Delta_A F_{\mu \nu} \right)\right],
   \tag{6.3}
   \label{eq:6.3}\\
    I_3 &= -\frac{e^2}{2} \int_{1/e\mathscr{F}}^{\infty} \dt e^{-tm^2} \TR \left( e^{-tD^2} F_{\mu \nu} \Delta_A F_{\mu \nu} \right).
   \tag{6.4}
   \label{eq:6.4}
\end{align}
At this point the choice of scaling parameter in \eqref{eq:6.2}-\eqref{eq:6.4} appears arbitrary. It is not for the following reasons.\\

(a) As remarked in Sec. IV, if the strong coupling behavior of $\det_\mathrm{{ren}}$ is to have anything to do with large amplitude variations of $F_{\mu \nu}$
then $e$ must appear in the combination $e\mathscr F$.\\

(b) The scaling parameter must be universal and not tied to any particular background field. As $m$ is always present in $\det_\mathrm{{ren}}$
it should be considered in the construction of a possible scaling parameter.\\

(c) The scaling parameter should result in the largest possible lower bound on $\Pi$ for $e \mathscr F  \gg m^2$.\\

(d) The lower bound should respect what is known about $\ln\det_\mathrm{ren}$'s mass dependence.\\

Based on (a)-(c) and the requirement that the scaling parameter have dimension (length)$^{-2}$ then possible scaling parameters have the
form $(e \mathscr F  )^a  m^b$, $2a + b = 2$, $a\ne0$. But only $a=1$, $b=0$ are allowed by requirement (d). To see why consider $I_1$  in \eqref{eq:6.2}. Following the result \eqref{eq:fourpointfive} for $e \mathscr F \gg m^2$,
\begin{align}
    I_1 = \frac{e^2 \norm{F}^2}{32 \pi^2} \ln \left( \frac{e \mathscr F}{m^2} \right) -\gamma + R,
   \tag{6.5}
   \label{eq:6.5}
\end{align}
where again $\gamma$ is Euler's constant and $0<|R|<m^2 /(e\mathscr F)$. The mass singularity in \eqref{eq:6.5} is induced by the on-shell charge
renormalization of $\ln\det_\mathrm{{ren}}$ in \eqref{eq:twopointone}, the starting point of this analysis. It is shown in Appendix F that for potentials $A_{\mu} \in \underset{r\ge 4- \epsilon}{\cap}  L^r ( \mathbb{R}^4)$,
$\epsilon>0$ and arbitrarily small, $\ln\det_\mathrm{{ren}}$ at $m=0$ is finite when it is renormalized off-shell.
Moreover, its $m=0$ limit is continuous. The restriction on $A_{\mu}$ excludes zero modes. Including them would cause lndet$_3$ to
diverge at $m=0$ as found in the results \eqref{eq:5.31} and \eqref{eq:5.39} that are independent of how $\ln\det_\mathrm{{ren}}$ is renormalized.

To define $\det_5$ in \eqref{eq:F1}, and therefore $\det_\mathrm{{ren}}$, it is sufficient to assume $A_{\mu} \in \underset{r>4}{\cap} L^r (\mathbb R ^4)$ \cite{7,31}. 
The charge renormalization term $\Pi$ depends only on $D^2$ and is therefore insensitive to zero modes. Without loss of generality we may assume here that $F_{\mu \nu} \in L^2$
and therefore that $A_{\mu} \in L^4$. This follows from the Sobolev inequality for gradients on $\mathbb R^4$ \cite{32}. Hence the restriction on $A_{\mu}$ in the preceding paragraph can be consistently assumed here.

When the first term in \eqref{eq:6.5} is combined with the mass singularity of $\ln\det_{SQED}$ in \eqref{eq:fourpointsix}, multiplied by $2$ as required
by \eqref{eq:threepointnine}, obtain
\begin{align}
    \ln\mbox{$\det{}_\mathrm{ren}$} \underset{m \to 0}{\sim} -\frac{e^2 \norm{F}^2}{48 \pi^2} \ln m^2 +\text{finite}.
   \tag{6.6}
   \label{eq:6.6}
\end{align}
The result in Appendix F allows us to state that this is the only divergent mass singularity of $\ln\det_\mathrm{{ren}}$ in the absence of zero
modes. If $\ln\det_\mathrm{{ren}}$ were subtracted off-shell by adding to \eqref{eq:twopointone}
the term
\begin{align}
    \frac{e^2 \norm{F}^2}{48 \pi^2} \int_0^{\infty} \frac{\dt}{t} \left( e^{-t\mu^2} - e^{-tm^2} \right) = \frac{e^2 \norm{F}^2}{48 \pi^2} \ln \left(  \frac{m^2}{\mu^2} \right),
   \tag{6.7}
   \label{eq:6.7}
\end{align}
then $\ln\det_\mathrm{{ren}}$ would be finite at $m=0$. This freedom to renormalize off-shell must be respected by the scaling parameter. Indeed, if
the scaling parameter $(e\mathscr F)^a   m^b$, $b\ne0$ were chosen in \eqref{eq:fourpointtwo} and \eqref{eq:6.2}-\eqref{eq:6.4} then \eqref{eq:6.6} would become
\begin{align}
   \ln\mbox{$\det{}_\mathrm{ren}$} \underset{m \to 0}{\sim} \left( \frac{b}{96} - \frac{1}{48} \right) \frac{e^2 \norm F^2}{\pi^2} \ln m^2 + \text{finite}.
   \tag{6.8}
   \label{eq:6.8}
\end{align}
This introduces a spurious $be^2  \norm F^2  \ln m^2  /96 \pi^2 $ mass singularity into $\ln\det_\mathrm{{ren}}$'s lower bound when it is renormalized off-shell using
\eqref{eq:6.7}. Therefore, the only acceptable scaling parameter for the strong coupling limit of $\Pi$ in \eqref{eq:6.1} and in $\det_\text{SQED}$ in \eqref{eq:fourpointtwo} is $e\mathscr F$. This further
justifies the choice of scaling parameter in Sec.IV.\\
 
\subsection{ Estimates}
Consider $I_2$  in \eqref{eq:6.3}. The trace in its last term can be put in the form $\TR(A^{\dagger} A)$ using the trace's cyclic property. So the last
term is not negative. Write out the trace term  in its original form and note that 
\begin{widetext}
  \begin{align}
    &\int_0^{1/e \mathscr F} \dt e^{-tm^2} \int \dfour x \dfour y e^{-tD^2}(x,y) F_{\mu \nu}(y) \Delta_A (y,x) F_{\mu \nu}(x) \notag\\
    &\le \int_0^{1/e\mathscr F} \dt e^{-tm^2}\int  \dfour x \left|\left(e^{-tD^2}   F_{\mu \nu} \Delta_A \right)(x) \right| \left|F_{\mu \nu}(x)\right| \notag \\
    &\le \int_0^{1/e\mathscr F} \dt e^{-tm^2} \int \dfour x \left( e^{-tP^2} \left| F_{\mu \nu} \right| \left| \Delta_A \right| \right)(x) \left|F_{\mu \nu} (x)\right| \notag \\
    &\le \int_0^{1/e\mathscr F} \dt e^{-tm^2}\int  \dfour x \left( e^{-tP^2} \left| F_{\mu \nu} \right|\Delta \right) (x) \left| F_{\mu \nu} (x) \right| \notag \\
    &=  \int_0^{1/e\mathscr F} \dt e^{-tm^2} \int \dfour x \dfour y e^{-tP^2}(x,y) \left| F_{\mu \nu}(x) \right| \Delta(y-x) \left| F_{\mu \nu} (x) \right|.
   \tag{6.9}
   \label{eq:6.9}
  \end{align}
\end{widetext}
To obtain these results we used the diamagnetic inequality of Simon \cite{12,33} to go from the second to the third line:
\begin{align}
    \left| (e^{-tD^2} f) (x) \right| \le \left( e^{-tP^2} \left|f \right| \right)(x). 
   \tag{6.10}
   \label{eq:6.10}
\end{align}
This holds for all $t>0$ and almost all $x \in \mathbb R^4$ and for potentials that are locally square integrable, as we are assuming. For more recent
comments on \eqref{eq:6.10} see \cite{34}. In addition we used Kato's inequality in the form given by \eqref{eq:A3} to go from the third to the fourth line.

Noting that
\begin{align}
    e^{-tP^2} (x,y) = \frac{1}{16 \pi^2 t^2} e^{- \left| x-y \right|^2/4t}, 
   \tag{6.11}
   \label{eq:6.11}
\end{align}
insertion of \eqref{eq:6.9} in \eqref{eq:6.3} gives
\begin{widetext}
\begin{align}
    I_2 \ge \frac{e^2}{32 \pi^2} \int_0^{1/e\mathscr F} \frac{\dt}{t} e^{-tm^2} \left( \norm F^2 -\frac{1}{t} \int \dfour x \dfour y \left| F_{\mu \nu}(x) \right| \Delta(x-y) e^{-(x-y)^2/4t} \left| F_{\mu \nu }(y)\right|  \right) .
   \tag{6.12}
   \label{6.12}
\end{align}
\end{widetext}
By Young's inequality in the form \cite{19}
\begin{align}
\left| \int \dfour x \dfour y f(x) g(x-y) h(y) \right| \le \norm f _p \norm g _q \norm h _r , 
\tag{6.13}
\label{eq:6.13}
\end{align}
where $1/p + 1/q + 1/r = 2$, $p,q,r\ge1$ and  $\norm f _p  = ( \int \dfour x \left|f(x)\right|^p )^{1/p}$, etc.,
\begin{align}
    I_2 \ge \frac{e^2 \norm F ^2}{32 \pi^2} \int_0^{1/e\mathscr F} \frac{\dt}{t} e^{-tm^2}\left( 1 - \frac{1}{t} \int \dfour x \Delta (x) e^{-x^2/4t}\right).
   \tag{6.14}
   \label{eq:6.14}
\end{align}
From $\Delta(x) = mK_1 (mx)/(4\pi^2 x)$ and integral 2.16.8.5 of \cite{35} get
\begin{align}
    I_2 \ge \frac{e^2 \norm F ^2}{32 \pi^2} \int_0^{1/e\mathscr F} \frac{\dt}{t} e^{-tm^2}\left[ 1 - m^2t e^{m^2t} \Gamma (-1, m^2t)\right],
   \tag{6.15}
   \label{eq:6.15}
\end{align}
where $\Gamma(-1, m^2 t)$ is the incomplete gamma function which we use in the form
\begin{align}
    \Gamma(-1, m^2t) = \frac{1}{m^2t}e^{-m^2t}- \int_{m^2t}^{\infty} \frac{\text d z}{z} e^{-z} .
   \tag{6.16}
   \label{eq:6.16}
\end{align}
Insertion of \eqref{eq:6.16} in \eqref{eq:6.15} and integrating by parts gives for $e\mathscr F   \gg m^2$
\begin{align}
    I_2 \ge \frac{e^2 \norm F ^2}{32 \pi^2} \left[ \frac{m^2}{e\mathscr F} \left( \ln \left( \frac{e \mathscr F}{m^2} \right) -\gamma + R \right)  +1 - e^{-m^2/e\mathscr F} \right],
   \tag{6.17}
   \label{eq:6.17}
\end{align}
with $\gamma$ and $R$ the same as in \eqref{eq:6.5}. Note that the lower bound in \eqref{eq:6.17} is finite at $m=0$ as it should be.

There are no ultraviolet divergences in $I_2$. The small $t$ dependence of the first term in \eqref{eq:6.3} is cancelled by the trace term, as was shown
in the above non-perturbative estimate. So it must be a general property of the trace term that
\begin{align}
    16 \TR \left( e^{-tD^2} F_{\mu \nu} \Delta_A F_{\mu \nu} \right) \underset{t \to 0}{\sim} \frac{\norm F ^2}{\pi^2 t} + \text{less singular in t}.
   \tag{6.18}
   \label{eq:6.18}
\end{align}
By inspection of \eqref{eq:6.3} we conclude that
\begin{align}
    \lim_{e \mathscr F \to \infty} \frac{I_2}{(e \mathscr F)^2 \ln \left( e \mathscr F \right)} = 0.
   \tag{6.19}
   \label{eq:6.19}
\end{align}

\def\textd{\text{d}}

Now consider $I_3$ in \eqref{eq:6.4}. As noted in the case of $I_2$ the trace is positive so that $I_3 \le  0$. Application of the inequality \eqref{eq:6.10} does
not lead to a satisfactory lower bound on $I_3$ . Namely, if it were saturated $I_3$ would cancel the large amplitude growth of $I_1$ in \eqref{eq:6.5},
resulting in a slow $O((e\mathscr F)^2)$ growth of $\Pi$ in \eqref{eq:6.1} and leading to the uninformative bound $\ln\det_\mathrm{{ren}} \ge -e^2 \norm F ^2 \ln (e\mathscr F/m^2)/96 \pi^2 + O\left((e\mathscr F)^2\right)$ following \eqref{eq:threepointnine} and \eqref{eq:fourpointsix}. 
We are confident that $\ln\det_\mathrm{{ren}}$ grows at least as fast as $ce^2 \norm F ^2 \ln(e\mathscr F)$, $c>0$, in the absence of zero mode supporting background fields.
This confidence is based on the result \cite{36} for the growth of $\ln\det_\mathrm{{ren}}$ for random, square-integrable, time-independent, non-zero mode supporting magnetic fields $\mathbf{B} (x)$ on $\mathbb R ^3$ ,
\begin{align}
\lim_{e \to \infty} \frac{\ln\mbox{$\det{}_\mathrm{ren}$}}{e^2 \ln e} = \frac{\norm{\mathbf{B}} ^2 T}{24 \pi^2},
\tag{6.20}
\label{eq:6.20}
\end{align}
where  $\norm{\mathbf B}^2  = \int \textd^3 x \mathbf B \cdot \mathbf B(x)$ and $T$ is the size of the time box. 
Therefore, our estimate of $I_3$  has to be more detailed than in the case of $I_2$. We claim that $\underset{e\to \infty}{\lim} I_3 /(e^2 \ln e) = 0$ for the class of fields considered here.\\

By summing over a complete set of scattering eigenstates $\ket{E, \alpha}$ of $D^2$, $I_3$  can be represented as
\begin{widetext}
  \begin{align}
    I_3 &= -\frac{e^2}{2} \sum_{\alpha, \beta} \int_{1/e \mathscr F}^{\infty} \textd t e^{-tm^2} \int_{0}^{\infty}\textd E e^{-tE} \int_{0}^{\infty} \textd E' \frac{\bra{E,\alpha} F_{\mu \nu} \ket{E', \beta} \bra{E', \beta} F_{\mu \nu} \ket{E,\alpha}}{E' + m^2} \notag \\
        &= -\frac{e^2}{2} \sum_{\alpha, \beta} \int_0^{\infty} \textd E \int_0^{\infty} \textd E' e^{-(E+m^2)/e\mathscr F} \frac{\left| \bra{E,\alpha} F_{\mu \nu} \ket{E', \beta} \right|^2}{(E+m^2)(E'+m^2)} ,
    \tag{6.21}
    \label{eq:6.21}
  \end{align} 
\end{widetext}
where $\alpha$  and $\beta$ are complete sets of angular momentum-like quantum numbers. Due to the above theorem on the $m=0$ limit of $\ln\det_\mathrm{{ren}}$ $I_3$ is finite at $m=0$. So whether $F_{\mu \nu}$  is long or short-ranged is irrelevant to the
growth of $I_3$  with $e$. Without loss of generality we may confine this discussion to fields with compact support. As $F_{\mu \nu}$   was assumed to be
differentiable in previous sections the compactly supported fields are assumed to rapidly and smoothly tend to zero in a narrow zone
near their boundries.  In addition we may assume rotational symmetry. Asymmetric, tangled fields will tend to lower the matrix elements
$\left|\bra{E,\alpha}F_{\mu \nu} \ket{E',\beta} \right|$. We will assume maximally symmetric $O(3)XO(2)$ fields to maximize $\left|I_3 \right|$.

For the potentials \eqref{eq:5.10} the equation for the radial part of the scattering states that satisfy  $D^2 \psi_{E,\alpha} = E \psi_{E,\alpha} $ is \cite{24}
\begin{widetext}
  \begin{align}
    \begin{split}
      \left( -\frac{\textd ^2}{\textd r^2} + \frac{(2j+1)^2-1/4}{r^2} + 4m_1 ea + e^2r^2a^2 \right) \phi_{Ejm_1}(r) = E \phi_{Ejm_1}(r), 
    \end{split}
    \tag{6.22}
    \label{eq:6.22}
  \end{align} 
\end{widetext}
where $\psi_{E,\alpha}(x) =  r^{-2j-3/2} \phi_{Ejm_1}(r) \mathscr D^j_{m_1 m_2} (x), r= \left| x\right|$, and the four-dimensional rotation matrices $\mathscr D_{m_1 m_2}^j$ are defined in Sec. V.B. 
Let $F_{\mu \nu}$ have range $R$. For $r>R$ the normalized wave function is, on setting the chiral anomaly equal to zero in \cite{24},
  \begin{align}
\begin{split}
     \phi_{Ejm_1}(r) &= \sqrt{\frac{r}{2}}J_{2j+1}(kr) \cos \delta_{j m_1} (k,e) \\
  &- \sqrt{\frac{r}{2}}Y_{2j+1}(kr) \sin \delta_{j m_1}(k,e),
    \end{split}
    \tag{6.23}
    \label{eq:6.23}
  \end{align} 
where  $\delta_{j m_1} (k,e)$ is the scattering phase shift in the indicated channel, $E=k^2$, and $Y_n$ is a Bessel function of the second kind.

We assumed in Sec. V.B that $a$ and $ra'$ are bounded functions of $r$. This will be assumed here. Therefore, any admissible $a$ maintains the
small distance behavior $\phi_{Ejm_1}\sim r^{2j+3/2}$ independent of $e$. What $\phi_{Ejm_1}$ does as $r\nearrow R$ is manifested in the exterior wave function \eqref{eq:6.23} through the
phase shifts. From \eqref{eq:6.22}, although $a$ descends rapidly to zero in a zone near $r=R$, it is evident from the $(era)^2$ term in \eqref{eq:6.22} that as
$e\to \infty$ there develops a high barrier at some point $r<R$ that blocks the entry of the exterior wave function \eqref{eq:6.23}, resulting in approximate
hard sphere scattering. This happens however rapidly $F_{\mu \nu}$ varies for $r<R$. So there is no reason why $F_{\mu \nu} = $ constant for $r<R$ and falling
rapidly to zero just before $r=R$ cannot be taken as representative of the general field case for the strong coupling estimate of $I_3$.

Accepting this, refer to \eqref{eq:5.10} and set $a(r) =  \lambda/R^2$  for $0 \le r \le R- \epsilon$ and $a(R ) = 0$. Then $F_{\mu \nu}  = 2 \lambda M_{\mu \nu}  /R^2$  for $0<r<R-\epsilon$.
The parameter $\lambda$ is related to the scaling parameter $\mathscr F$ by $\mathscr F^2 = F_{\mu \nu}^2 = 16 \lambda^2/R^4$ since $M_{\mu \nu}^2 = 4$. Then
  \begin{align}
\begin{split}
     &\bra{E j m_1} F_{\mu \nu} \ket{E' j' m_1'} \\
    &= \frac{4 \pi^2 \lambda M_{\mu \nu}}{2j+1} \delta_{j j'} \delta_{m_1 m_1'} \int_0^R \textd r \phi_{E j m_1} \phi_{E' j m_1},
   \end{split}   
 \tag{6.24}
    \label{eq:6.24}
  \end{align}                                                 
where we have taken the limit $\epsilon  = 0$ on the right-hand side of \eqref{eq:6.24}. As shown below it follows from \eqref{eq:6.22} that
  \begin{align}
    \begin{split}
     &\left( \phi_{E'jm_1} \phi'_{Ejm_1} - \phi_{Ejm_1}\phi'_{E'jm_1}\right)(R) \\
     &= (E' - E) \int_0^R \textd r \phi_{Ejm_1} \phi_{E'jm_1}.
     \end{split}
    \tag{6.25}
    \label{eq:6.25}
  \end{align} 
Then \eqref{eq:6.24} and \eqref{eq:6.25} combined with \eqref{eq:6.21} give
\begin{widetext}
  \begin{align}
\begin{split}
     I_3 =& -2\pi^4(e\mathscr F)^2 \int_0^{\infty} \textd E \int_0^{\infty} \textd E' e^{-(E+m^2)/e\mathscr F} \\
     &\times\sum_{j=0,1/2,..}^{\infty} \frac{1}{(2j+1)^2} \sum_{m_1,m_2 = -j}^{j}  
     \frac{\left[ \left( \phi_{E'jm_1} \phi'_{Ejm_1} - \phi_{Ejm_1} \phi'_{E'jm_1} \right)(R) \right]^2}{(E+m^2)(E'-E)^2(E'+m^2)}.
\end {split}     
\tag{6.26}
    \label{eq:6.26}
  \end{align} 
\end{widetext}

To obtain \eqref{eq:6.25} from the assumed behavior of $F_{\mu \nu}$  multiply \eqref{eq:6.22} at energy $E$ by $\phi_{E'jm_1}(r) F_{\mu \nu}(r)$, subtract the result with $E\leftrightarrow E'$
and integrate by parts over the interval $0 \le r \le R$. Since $F_{\mu \nu}(R)=0$ and $\phi_{Ejm_1}(0)=0$ this gives
\begin{widetext}
  \begin{align}
     \int_{R-\epsilon}^R \textd& r \left( \phi_{E'jm_1} \phi'_{Ejm_1} - \phi_{Ejm_1} \phi'_{E'jm_1} \right) \frac{\textd F_{\mu \nu}(r)}{\textd r} \notag \\
        =&\left( E- E' \right) \left[ \frac{2 \lambda M_{\mu \nu}}{R^2} \int_0^{R-\epsilon} \textd r \phi_{Ejm_1} \phi_{E'jm_1} + \int_{R-\epsilon}^R \textd r \phi_{Ejm_1} \phi_{E'jm_1} F_{\mu \nu}(r)  \right].
   \tag{6.27}
   \label{eq:6.27}
 \end{align} 
\end{widetext}
Assuming $\epsilon/R \ll 1$ and noting that $\int_{R-\epsilon}^R \textd r F'_{\mu \nu}(r) = - F_{\mu \nu}(R-\epsilon) = -\frac{2\lambda M_{\mu \nu}}{R^2}$, \eqref{eq:6.25} follows after letting $\epsilon \to 0$.

The phase shifts required to calculate $I_3$ are obtained as follows.
Set $a = \lambda /R^2$ in \eqref{eq:6.22} and let, omitting subscripts,
  \begin{align}
     \phi (r) = r^{2j + 3/2} f(r) e^{-\lambda e r^2 /2R^2}.
    \tag{6.28}
    \label{eq:6.28}
  \end{align} 
Then
  \begin{align}
\begin{split}
     &f'' + \left( \frac{4j+3}{r} - \frac{2 \lambda e r}{R^2} \right) f' + \left[ k^2 - \frac{4\lambda e}{R^2}(j+m_1+1) \right] f \\
      &= 0.
\end{split}
    \tag{6.29}
    \label{eq:6.29}
  \end{align} 
The solution of \eqref{eq:6.29} regular at the origin is the confluent hypergeometric function
  \begin{align}
     f(r) = M\left( j+m_1+1 - \frac{(kR)^2}{4\lambda e}, 2j+2, \frac{\lambda e r^2}{R^2} \right),
    \tag{6.30}
    \label{eq:6.30}
  \end{align} 
following the notation of \cite{37}. Joining \eqref{eq:6.23} with \eqref{eq:6.28} at $r=R$ gives
  \begin{align}
\begin{split}
     &\tan \delta_{j m_1} (k,\lambda \text{e}) = \\
    &\frac{(\gamma -1/2)J_{2j+1}(kR)-kR J'_{2j+1}(kR)}{(\gamma-1/2)Y_{2j+1}(kR)-kR Y'_{2j+1}(kR)},
    \end{split}   
    \tag{6.31}
    \label{eq:6.31}
  \end{align} 
where $\gamma = (r \phi'  /\phi)_R$. Eqs. \eqref{eq:6.28}, \eqref{eq:6.30} and Eq.(13.4.8) in \cite{37} for $\textd M(a,b,z)/\textd z$ give
  \begin{align}
     \gamma = 2j + \frac{3}{2} - \lambda e + \frac{2\lambda e a}{b} \frac{M(a+1, b+1, \lambda e)}{M(a,b,\lambda e)},
    \tag{6.32}
    \label{eq:6.32}
  \end{align}                                                                  
where $a = j + m_1  + 1 - (kR)^2 /(4 \lambda  e), b = 2j + 2$. There are several cases. For $j < \lambda e \gg 1$, fixed $k$,
  \begin{align}
    \begin{split}
    \gamma = &\lambda e + 2m_1 -\frac{1}{2} - \frac{(kR)^2}{2e\lambda} \\
    &+ O\left( \frac{j^2}{\lambda e}, \frac{j(kR)^2}{(\lambda e)^2}, \frac{(kR)^4}{(\lambda e)^3} \right).
    \end{split}
    \tag{6.33}
    \label{eq:6.33}
  \end{align} 
For $j > \lambda  e \gg 1$, fixed k,
  \begin{align}
     \gamma = 2j + \frac{m_1}{j} \lambda e - \frac{(kR)^2}{4j} + O\left( \frac{\lambda e}{j^2}, \frac{m_1 \lambda e}{j^2}, \frac{(kR)^2}{j^2}\right),
    \tag{6.34}
    \label{eq:6.34}
  \end{align} 
and for $kR \to \infty$, fixed $j$, $\lambda e$,
  \begin{align}
     \gamma = - \left[ kR + O\left( \frac{1}{kR} \right) \right] \frac{J_{2j+2}(kR)}{J_{2j+1}(kR)} + 2j - \lambda e + 3/2.
    \tag{6.35}
    \label{eq:6.35}
  \end{align}                                                                  
These results are obtained using the asymptotic expansions of $M(a,b,z)$ for large $a, b, z$ in \cite{37,38}. Following \eqref{eq:6.35} the phase shifts vanish
at high energy as $\tan \delta \sim (e \lambda /kR)\cos^2 (kR-\left(j+ \frac{1}{2} \right) \pi  - \pi /4)$.

In order to estimate $I_3$  for $e\mathscr{F}\rightarrow\infty$ it is convenient to divide the range of the $kR$, $k'R$ integrations in \eqref{eq:6.26} into $[0,2)$, $[2, 2\sqrt{ e \mathscr F R^2} )$, 
$[2\sqrt{ e \mathscr F R^2} , 2(e \mathscr F  R^2 )^{1-\epsilon}), [2e \mathscr F  R^2 , \infty ]$ and the special case
$kR, k'R = O(e \mathscr F R^2 )^{1-\epsilon}$, where $0< \epsilon \ll 1$. To accommodate the joining conditions \eqref{eq:6.33}-\eqref{eq:6.35} the range of $j$ also has to be partitioned. 
It is essential not to interchange the large $e\mathscr F$ limit with the sum over $j$. We find that the dominant contributions to \eqref{eq:6.26} come from
$0\le j\le \sqrt{ e \mathscr F  R^2} , 2 \le kR \lesssim O( \sqrt{e\mathscr F  R^2} )$ and $2 \le k'R \le \infty$. There are many cases
to consider; we outline here a representative case to indicate how the estimates are done.

 Consider the contribution to \eqref{eq:6.26} given by
   \begin{widetext}
   \begin{align}
     I \equiv - 8 \pi^4 (e \mathscr F R^2)^2 \sum_{j=0}^{\sqrt{e\mathscr F R^2}}\sum_{m_1, m_2 = -j}^{j} \frac{1}{(2j+1)^2} &\int_{2j+1}^{2\sqrt{e \mathscr F R^2}} \frac{\textd (kR)}{kR} e^{-\frac{k^2}{e\mathscr F}}\int_{2\sqrt{e\mathscr F R^2}}^{2(e \mathscr F R^2)^{1-\epsilon}} \frac{\textd (k'R)}{k'R}  \notag \\
      & \times \frac{\left[ \left( \phi_{E'jm_1} \phi'_{Ejm_1} - \phi_{Ejm_1} \phi'_{E'jm_1} \right)(R) \right]^2}{R^4(k'^2-k^2)^2},
    \tag{6.36}
    \label{eq:6.36}
  \end{align}
  \end{widetext}
where we have noted above that we can set $m=0$. For the range of $kR$, $k'R$ and $j$ in \eqref{eq:6.36} joining condition \eqref{eq:6.33} applies. From \eqref{eq:6.23}, \eqref{eq:6.31}
and \eqref{eq:6.33} obtain
  \begin{widetext}
  \begin{align}
     \frac{(\phi_{E'}\phi'_{E} - \phi_E \phi'_{E'})(R)}{k'^2-k^2} \underset{\lambda e \gg 1}{\sim} \frac{R^2}{2\pi^2(\lambda e)^3} \left[ \left(J_{2j+1}(kR) - \frac{kR}{\gamma -1/2} J'_{2j+1}(kR) \right)^2 + \left( Y_{2j+1} (kR) - \frac{kR}{\gamma -1/2} Y'_{2j+1}(kR)\right)^2 \right]^{-1/2} \notag \\
     \times \Big[ k \to k' \Big]^{-1/2} \sim \frac{R^2}{2\pi^2 (\lambda e)^3}\left(J_{2j+1}^2(kR) + Y^2_{2j+1}(kR) \right)^{-1/2} \left(J_{2j+1}^2(k'R) + Y_{2j+1}^2(k'R) \right)^{-1/2}.
    \tag{6.37}
    \label{eq:6.37}
  \end{align}
  \end{widetext}
Hence,
  \begin{widetext}
  \begin{align}
     I \underset{e\lambda \gg 1}{\sim} - \frac{8192}{(e\mathscr F R^2)^4} \sum_{j=0}^{\sqrt{e\mathscr F R^2}} \int_{2j+1}^{2\sqrt{e\mathscr F R^2}} \frac{\textd (kR)}{kR} e^{-k^2/e\mathscr F} \frac{1}{J_{2j+1}^2(kR) + Y_{2j+1}^2(kR)} \notag \\ \times \int_{2\sqrt{e\mathscr F R^2}}^{2(2\mathscr F R^2)^{1-\epsilon}} \frac{\textd (k'R)}{k'R} \frac{1}{J_{2j+1}^2(k'R) + Y_{2j+1}^2(k'R)},
    \tag{6.38}
    \label{eq:6.38}
  \end{align} 
  \end{widetext}
where the sums over $m_1$ and $m_2$ have been taken. To estimate \eqref{eq:6.38} use Watson's inequality (Eq.(1), Sec.13.74 of \cite{39})
  \begin{align}
     \frac{2}{\pi x} < J_n^2(x) + Y_n^2(x) < \frac{2}{\pi}(x^2-n^2)^{-1/2},
    \tag{6.39}
    \label{eq:6.39}
  \end{align} 
for $x \ge n \ge 1/2$. This is used repeatedly in our estimates. An easy calculation gives
  \begin{align}
     I \underset{e\lambda \gg 1}{=} O\left( -(e \mathscr F R^2)^{-2-\epsilon} \right),
    \tag{6.40}
    \label{eq:6.40}
  \end{align}                                                                  
with $0< \epsilon \ll 1$. The remaining contributions to $I_3$  give
  \begin{align}
     I_3 \underset{e\lambda \gg 1}{=} O\left( -(e \mathscr F R^2)^{-2} \right),
    \tag{6.41}
    \label{eq:6.41}
  \end{align}                                                               
or smaller as in \eqref{eq:6.40}. The dominant estimate in \eqref{eq:6.41} comes from the intervals $0 \le j \le \sqrt{  e \mathscr F  R^2  }, 2j+1 \le kR \le O((\sqrt{ e \mathscr F R^2  })$,
$O(e\mathscr F  R^2 ) \le k'R \le \infty$.

 We have given reasons above why this calculation of the large amplitude growth of $I_3$ is representative. In view of \eqref{eq:6.41} we are
confident that
  \begin{align}
     \lim_{e\mathscr F \to \infty} \frac{I_3}{(e\mathscr F)^2 \ln (e\mathscr F)} = 0,
    \tag{6.42}
    \label{eq:6.42}
  \end{align} 
for all admissible random fields. Combining \eqref{eq:6.1}, \eqref{eq:6.5}, \eqref{eq:6.19} and \eqref{eq:6.42} we obtain for large amplitude variations of admissible
random fields $F_{\mu \nu}$
  \begin{align}
     \Pi \underset{e\mathscr F \to \infty}{=} \frac{e^2 \norm F ^2}{32 \pi^2} \ln\left( \frac{e \mathscr F}{m^2} \right) + R_1,
    \tag{6.43}
    \label{eq:6.43}
  \end{align} 
with $\underset{e\mathscr F \to \infty}{\lim} R_1 /[(e \mathscr F )^2 \ln(e \mathscr F)] = 0$. The term "admissible random field" is discussed in Sec. VII.

\subsection{Summary}
In the absence of zero mode supporting random background fields \eqref{eq:threepointnine}, \eqref{eq:fourpointsix}, \eqref{eq:5.24} and \eqref{eq:6.43} give the final result
  \begin{align}
     \text{lndet}_{ren} \underset{e \mathscr F \to \infty}{\ge} \frac{1}{48 \pi^2} e^2 \norm F ^2  \ln \left( e \mathscr F / m^2 \right) + R_2,
    \tag{6.44}
    \label{eq:6.44}
  \end{align} 
with $R_2$'s growth bounded as $R_1$'s above. The lnm$^2$   contribution to \eqref{eq:6.44} is due to on-shell charge renormalization. For off-shell renormalization $m^2$ is
replaced with a subtaction parameter $\mu^2$ as discussed in Sec. A above. 

If zero mode supporting background fields are included and all of
the zero modes have the same chirality then by \eqref{eq:threepointnine}, \eqref{eq:fourpointsix}, \eqref{eq:5.39}(an equality in this case) and \eqref{eq:6.43}.
\begin{align} 
\begin{split}
     \ln\mbox{$\det{}_\mathrm{ren}$} \underset{e\mathscr F \to \infty}{\ge}& - \frac{1}{16 \pi^2} \left| \int \textd ^4 x ^* F_{\mu \nu} F_{\mu \nu} \right| e^2 \\
    &\times\ln(\frac{e \mathscr F}{m^2}) + \frac{1}{48 \pi^2} e^2 \norm F ^2 \ln(\frac{e \mathscr F}{m^2}) + R_3,
    \end{split}
    \tag{6.45}
    \label{eq:6.45}
  \end{align}
with $R_3$ bounded as $R_1$ and $R_2$ above. Recall that $ \int \textd ^4 x^*  F_{\mu \nu}  F_{\mu \nu} /16\pi^2$ is the
chiral anomaly. 

If the zero modes supported by $A_{\mu}$ have both positive and negative chirality there is no counting theorem and \eqref{eq:6.45} is replaced with, following \eqref{eq:5.31} and \eqref{eq:5.32},
  \begin{align}
\begin{split}
	\ln\mbox{$\det{}_\mathrm{ren}$} \underset{e\mathscr F \to \infty}{\ge}& - \left( \# \text{zero modes supported by }A_{\mu} \right)\\
      &\times \ln(\frac{e \mathscr F}{m^2}) + \frac{1}{48 \pi^2}e^2 \norm F ^2 \ln(\frac{e \mathscr F}{m^2}) + R_4. 
    \end{split}
    \tag{6.46}
    \label{eq:6.46}
  \end{align}
The number of zero modes grows at least as fast as $e^2$ following \eqref{eq:5.37}, provided the chiral anomaly is non-zero.. If they grow as $e^2$ or less then $\underset{e \mathscr F \to \infty}{\lim}  R_4 /[(e \mathscr F )^2 \ln(e \mathscr F )] = 0$.

Known 4D Abelian zero modes require $F_{\mu \nu} \not\in L^2$. So the $\norm F ^2$ terms in \eqref{eq:6.45} and \eqref{eq:6.46} need a volume cutoff that will be discussed in Sec.VII.
Assuming in this section that $F_{\mu \nu} \in L^2$ served its purpose to obtain the structure of the charge renormalization term's large field amplitude contribution to $\ln\det_\mathrm{{ren}}$.

An assumption underlying \eqref{eq:6.46} is that all admissible 4D Abelian zero mode supporting potentials have a $1/|x|$ falloff as $|x|\rightarrow \infty$. If there were zero mode supporting potentials whose falloff is faster than $1/|x|$ the associated chiral anomaly would vanish since $^*  F_{\mu \nu}  F_{\mu \nu}=\partial_{\alpha}(\epsilon _{\alpha\beta\mu\nu}\text{A}_{\beta}F_{\mu\nu}) $. 
The vanishing of the right-hand side of \eqref{eq:5.37} implies $n_+=n_-$. Without being able to place a lower bound on the number of zero modes \eqref{eq:6.46} loses its predictive power in this case. A 4D Abelian vanishing theorem stating that all normalizable zero modes have either positive or negative chirality, as in QCD$_4$, needs to be either proved or falsified by a counterexample.

Further discussion of \eqref{eq:6.44}-\eqref{eq:6.46} appears at the end of Sec. VII.

\section{Regularization}

In principle det$_{ren}$ can be calculated as an explicit function of $F_{\mu \nu}$ before inserting it into the functional integral \eqref{eq:twopointfive}. The
input potentials must correspond to random potentials supported by d$\mu_0(A)$. It is generally accepted that these belong to $\mathscr S '(\mathbb{R}^4)$, the
space of tempered distributions. This is the first requirement.

     Throughout we have assumed smooth potentials, including zero mode supporting potentials $A_{\mu} (x)$ with a $1/|x|$ falloff for $|x| \to \infty$.
In Sec.VA it was assumed that $F_{\mu \nu} \in \underset{r>2}{\cap}L^r(\mathbb R^4)$ which we noted may be
too strong a condition.  The $L^p (\mathbb R^4)$ Sobolev inequality $\|\nabla f\|_p \ge K\|f\|_q$,
where $K$ is a constant and $q=4p/(4-p),1<p<4$ \cite{32}, implies $A_{\mu} \in \underset{r>4}{\cap} L^r ( \mathbb R^4 )$
when $A_{\mu}$ is once differentiable and $F_{\mu \nu} \in \overset{<4}{\underset{>2}{\cap}} L^r(\mathbb R^4)$. This condition
on $A_{\mu}$ and the weaker condition on $F_{\mu\nu}$ are sufficient to define det$_5$  in \eqref{eq:F1} to ensure that $\ln\det_\mathrm{{ren}}$ is defined when $m\neq 0$ \cite{7,31}. These assumptions constitute the second requirement.
 
The final requirement is that an ultraviolet cutoff mechanism be introduced.

These three requirements can be satisfied by calculating $\ln\det_\mathrm{{ren}}$ in terms of the potentials
\begin{align}
A_{\mu}^{\Lambda} (x) = \int \text{d}^4 y f_{\Lambda}(x-y) A_{\mu}(y),
\tag{7.1}
\label{eq:7.1}
\end{align}
     where $A_{\mu} \in \mathscr S'(\mathbb R^4)$ and $f_{\Lambda} \in \mathscr S(\mathbb R^4)$, the space of functions of rapid
decrease. Then $A_{\mu}^{\Lambda} \in \text{C}^{\infty}$. Besides smoothing $A_{\mu}$, \eqref{eq:7.1} also introduces a
sequence of ultraviolet cutoffs. Thus, from \eqref{eq:twopointthree} conclude that
\begin{align}
\int \text{d} \mu _0 (A) A_{\mu}^{\Lambda}(x) A_{\nu}^{\Lambda}(y) = D_{\mu \nu}^{\Lambda}(x-y),
\tag{7.2}
\label{eq:7.2}
\end{align}
where the Fourier transform of the regularized free photon propagator in a fixed gauge is $\hat D_{\mu \nu}(k)|\hat f_{\Lambda} (k)|^2$ with $\hat{f}_{\Lambda} \in \text{C}_0^{\infty}$, the space of C$^{\infty}$
functions with compact support. For example, one might choose $\hat f_{\Lambda}   = 1$, $k^2\le \Lambda ^2$ and $\hat f_{\Lambda} = 0, k^2 \ge n \Lambda^2 , n>1$.

We note that if $A_{\mu}$  is a zero mode supporting potential then so is $A_{\mu}^{\Lambda}$. Thus, if $A_{\mu}$ has a $1/|x|$ falloff then so does $A_{\mu}^{\Lambda}$ . 
This follows since the small-$p$ dependence of their Fourier transforms, and hence their large-$x$ dependence, are the same when $\hat f_{\Lambda}$   is chosen as above;
chirality is preserved. Other mappings with the convolution in \eqref{eq:7.1} can be followed with Young's inequality in the form \eqref{eq:A7} with $s=1$; the above conditions on $A_\mu$ and $F_{\mu\nu}$ are preserved.

Summarizing, we are instructed to replace all potentials and fields in this analysis with the smoothed potentials $A_{\mu}^{\Lambda}$ and fields $F_{\mu \nu}^{\Lambda} = \partial_{\mu} A_{\nu}^{\Lambda} - \partial_{\nu}A_{\mu}^{\Lambda}$, including the general representation \eqref{eq:twopointfive}. This allows the assumed restrictions on $A_{\mu}$ and $F_{\mu\nu}$ leading to \eqref{eq:6.44}-\eqref{eq:6.46} to be transferred to $A_{\mu}^{\Lambda}$ and $F_{\mu\nu}^{\Lambda}$  while keeping the underlying rough potentials $A_{\mu}$ in place.

The measure d$\mu_0(A)$ is not modified.
The substitution of $A_\mu^\Lambda$  for $A_\mu$  does not affect the analysis of Secs.V A-D. In particular, in Sec.V B where use is made of \eqref{eq:5.10}
we have
\begin{align}
\begin{split}
\hat{A}_\mu(k)&=M_{\mu\nu} \int \text{d}^4x \text{e} ^{-ikx} x_\nu a (r)\\
&= i M_{\mu\nu} \partial_\nu \hat{a}(|k|).
\end {split}
\tag{7.3}
\label{eq:7.3}
\end{align}
Then
\begin{align}
\begin{split}
A_\mu^\Lambda(x)&= \int \text{d}^4y f_\Lambda(x-y) A_\mu(y)\\
&= (a_\Lambda(r)+ \text{h}_\Lambda(r)) M_{\mu\nu} x_\nu,
\end {split}
\tag{7.4}
\label{eq:7.4}
\end{align}
where
\begin{align}
 a_\Lambda(r)=\int \frac{\text{d}^4k}{(2\pi)^4}\text{e}^{ikx}\hat{a}(|k|)\hat{f}_\Lambda(|k|),
\tag{7.5}
\label{eq:7.5}
\end{align}
\begin{align}
 h_\Lambda(r)x_\nu=-i \int \frac{\text{d}^4k}{(2\pi)^4} \text{e}^{ikx}\hat{a}(|k|) \partial_\nu \hat{f}_\Lambda(|k|).
\tag{7.6}
\label{eq:7.6}
\end{align}
If $A_\mu$  supports a zero mode then $a_\Lambda(r)\underset{r \rightarrow \infty}{\sim}\nu/r^2 $ since $\hat{f}_\Lambda(|k|)=1$ for $k^2 \le \Lambda^2$. 
Hence, the only result of substituting $A_\mu^\Lambda$  for $A_\mu$  is to replace $a$ with $a_\Lambda + h_\Lambda$.

In Sec.V E the profile function $a(r)$ in \eqref{eq:5.40} has a discontinuous second derivative at r=R.
So $a(r)$ for $r\le R$ would have to be smoothed to accommodate a reqularized potential. This does not in any way modify the conclusion of Sec.V E, namely that the
formalism of Secs.V C and D can be implemented.

In Sec.VI B we can not choose $F_{\mu\nu}^\Lambda \in  \text{C}_0^\infty$  as we did for $F_{\mu\nu}$. For suppose $F_{\mu\nu}^\Lambda \in \text{C}_0^\infty$.
Then $\hat{F}_{\mu\nu}^\Lambda(k)$ is an entire analytic function of $k_\mu$  \cite{40}. Therefore, we cannot set  $\hat{F}_{\mu\nu}^\Lambda(k)=\hat{f}_\Lambda(|k|) \hat{F}_{\mu\nu}(k)$ since $\hat{f}_\Lambda (|k|)$ is not an entire analytic function of $|k|$. Nevertheless, $F_{\mu\nu}^\Lambda(x) = f_\Lambda \\  ^* F_{\mu\nu}(x)$ is a polynomial bounded C$^\infty$ function by Theorem IX.4 in \cite{40}.  We are now free to choose a $F_{\mu\nu} \in \mathscr S'$ to make $F_{\mu\nu}^\Lambda(x)$ fall off arbitrarily rapidly for $|x|>R$. So $F_{\mu\nu}^\Lambda$ can be chosen arbitrarily close to a compactly supported field. This should not change our conclusion \eqref{eq:6.42} about the bound on $I_3$  for $e>>1$.

Finally, a volume cutoff must be introduced in $\text{det}_\text{ren}$ -and only $\text{det}_\text{ren}$- in order to regularize the vacuum-vacuum amplitude Z in \eqref{eq:twopointfour}. As $\text{det}_\text{ren}$  is gauge invariant this can be done by letting $F_{\mu\nu}^\Lambda \rightarrow \text{g}F_{\mu\nu}^\Lambda$ , where g is a space cutoff such as $\text{g} \in \text{C}_0^\infty$ or $\text{g}=\chi _{\Gamma}$, the characteristic function of a bounded region $\Gamma \subset \mathbb R^4$. This way of introducing g preserves the gauge invariance of $\text{det}_\text{ren}$   .

The regularization procedure used here is a generalization of that used in the two-dimensional Yukawa model \cite{41}. The main conclusions in this paper obtained without regulators remain valid. Thus, in \eqref{eq:6.44}-\eqref{eq:6.46} it is only required to replace $F_{\mu\nu}$  with $\text{g}F_{\mu\nu}^\Lambda$, which is a
special case of the general substitution $\text{det}_\text{ren}(F_{\mu\nu})\rightarrow\text{det}_\text{ren}(\text{g}F_{\mu\nu}^\Lambda)$. $\mathscr F$ is the amplitude of $F_{\mu\nu}^{\Lambda}$ whose scale is set by the amplitude of the underlying potential $A_\mu \in \mathscr S'$.
 It does not matter when the regulators are introduced as long as they are in place when $\text{det}_\text{ren}$  is inserted into \eqref{eq:twopointfive} .   

{\it{Interpretation of \eqref{eq:6.44}-\eqref{eq:6.46}}}: Each term in representation \eqref{eq:threepointnine} for $\det_\mathrm{ren}$ is gauge invariant and ultraviolet finite. Therefore, each term is independent of the others. It is noted in \eqref{eq:6.44}-\eqref{eq:6.46}, with $F_{\mu \nu}$ replaced by $F_{\mu \nu}^{\Lambda}$ before introducing $g$, that $F_{\mu \nu}^{\Lambda}$ must be square integrable. Within the class of potentials with falloff at infinity those that support a zero mode decrease as $1/|x|$ as far as presently known. This is incompatible with $F_{\mu\nu}^{\Lambda} \in \text{L}^2$. The
terms in \eqref{eq:6.44}-\eqref{eq:6.46} depending on $||F^{\Lambda}||^2$ come from the first and third terms of \eqref{eq:threepointnine}. These terms were dealt with in Secs. IV and VI where it was assumed that $F_{\mu\nu}^{\Lambda} \in \underset{r\ge 2}{\cap} L^r$. Zero modes reside solely in the second term of \eqref{eq:threepointnine}. As shown in Sec. V it can be defined for $F_{\mu\nu}^{\Lambda} \in \underset{r>2}{\cap} L^r$. So the two terms in \eqref{eq:6.45} and \eqref{eq:6.46} are separately defined, each subject to its foregoing field restriction.

 To regulate Z in \eqref{eq:twopointfour} a volume cutoff is inserted into $\det_\mathrm{ren}$ as described above. When zero mode supporting potentials are introduced into $\det_\mathrm{ren}$ by the Maxwell measure $d{\mu}_0(A)$ the terms depending on  $||F^{\Lambda}||^2$ now remain finite. Therefore, the interpretation of \eqref{eq:6.44}-\eqref{eq:6.46} is that they represent the asymptotic {\it{form}} of $\det_\mathrm{ren}$ before volume cutoffs are introduced.

For \eqref{eq:6.44}-\eqref{eq:6.46} to be relevant the unregularized random connections $A_{\mu}$, including their assumed falloff at infinity, should have $\mu_0$ measure one. As far as the author knows all known results for the growth at infinity of a set of random fields with measure one are for a Gaussian process whose covariance corresponds to a massive scalar field (see, for example, \cite{52, 53}). The covariance (\ref{eq:twopointthree}) in a general covariant gauge does not include an infrared cutoff photon mass as none is required. To the author's knowledge, then, the behavior at infinity of a set of random Euclidean QED$_4$ connections with $\mu_0$ measure one is still not settled. 

\section{Conclusion}
    Representations \eqref{eq:twopointsix} and \eqref{eq:threepointnine} for the Euclidean fermion determinant in QED, $\ln\det_\mathrm{{ren}}$, have been obtained that reflect
its competing magnetic properties of diamagnetism and paramagnetism.
This way of viewing $\ln\det_\mathrm{{ren}}$ arises since in Euclidean space $F_{\mu \nu}(x)$ may be regarded as a static, four-dimensional magnetic field. This
decomposition of $\ln\det_\mathrm{{ren}}$ has the advantage of simplifying its strong coupling, large field amplitude analysis for a class of random
potentials/fields. The analysis is made possible by a number of theorems developed in the 1970s and 80s that are applicable
to field-theoretic operators in the presence of external gauge fields.

The main results are summarized by \eqref{eq:6.44}-\eqref{eq:6.46} and are interpreted at the end of Sec. VII. Result \eqref{eq:6.44} for the fast growth of $\ln\det_\mathrm{{ren}}$ for large
field variations raises doubt on whether it is integrable with any Gaussian measure whose support does not include zero mode supporting potentials. Results \eqref{eq:6.45} and \eqref{eq:6.46} indicate that the growth of $\ln\det_\mathrm{{ren}}$ is slowed down or stopped by including zero mode supporting
potentials in the Gaussian measure d$\mu_0(A)$ introduced in Sec.II. This is \textit{prima face} evidence that zero mode supporting potentials are
necessary for the non-perturbative quantization of QED. See \cite{54} for an earlier discussion of the non-perturbative quantization of QED.

     Refer back to one of the electroweak fermion determinants such as the first one in \eqref{eq:onepointone}. Suppose after being properly defined its
large amplitude Maxwell field variation coincides with that of $\ln\det_\mathrm{{ren}}$. Then \eqref{eq:6.45} and \eqref{eq:6.46} provide \textit{prima face} evidence that the
non-perturbative quantization of the electroweak model also requires its Maxwell Gaussian measure to have support from zero mode supporting
potentials. This assumes that the Maxwell field integration follows next after integrating out the fermion degrees of freedom.

     Given such Gaussian measures are they such that no measurable subset of potentials results in the fast growing charge renormalization
term in \eqref{eq:6.45} and \eqref{eq:6.46} becoming dominant? This is entering unknown territory that needs to be explored.

If the QED determinant grows faster than a quadratic in the Maxwell field for a measurable set of fields then there may be a connection between this and the photon propagator's Landau pole \cite{5,56}. The precise connection, if any, remains to be worked out.

It might be objected that the non-perturbative quantization of the electroweak model is irrelevant since perturbative expansions appear to be adequate at presently available energies. This opinion neglects the fact that the electroweak model is not asymptotically free. At some point the model's non-perturbative content will be required.

The author wishes to acknowledge helpful correspondence with Erhard Seiler.

\appendix

\def\FD{\mathinner{\Delta_A^{1/2}}}
\def\I3{\mathinner{\mathscr I_3}}
\def\R4{\mathinner{\mathbb R^4}}
\def\TR{\mathinner{\mathrm{Tr}}}

\section{$\FD\sigma F\FD$}
It is claimed that $\FD\sigma F\FD$ belongs to the trace ideal $\I3$ for $F_{\mu\nu} \in \cap_{q>2}L^q (\R4  )$.
The trace ideal $\mathscr I_p (1\le p<\infty)$ is defined as  those compact operators $A$ with  $\norm{A}^p_p = \TR((A^\dagger A)^{p/2}  )< \infty$.
General  properties of  $\mathscr I_p$  spaces used here may be found in \cite{8,9,10}.
To  simplify notation we set $e=1$ in this appendix.

To decide whether $\FD\sigma F\FD\in\I3$ it suffices to deal with $\FD\abs{F}\FD$ ($F_{\mu\nu}^2=\abs{F}^2$)  since $\sigma F/\abs{F}$ is unitary.
Then $\FD\abs{F}\FD\in\I3$ if $\abs{F}^{1/2}\FD\in\mathscr I_6$ since by H\"older's inequality for $\mathscr I_p$ spaces
\begin{align}
    \norm{\FD\abs F\FD}_3\le\norm{\FD\abs F^{1/2}}_6\norm{\abs F^{1/2}\FD}_6.
    \label{eq:A1}
\end{align}
If $\abs F^{1/2}\FD\in\mathscr I_6$ then so does its adjoint $\FD\abs F^{1/2}$ by the general  properties of $\mathscr I_p$ spaces.
Then
\begin{align}
    \begin{split}
        \norm{\abs F^{1/2}\FD}_6^6 &= \TR\left( \FD\abs F\Delta_A\abs F\Delta_A\abs F\FD \right)\\
        &\le \TR \left( \Delta^{1/2}\abs F\Delta\abs F\Delta\abs F\Delta^{1/2} \right)\\
        &=  \norm{\abs F^{1/2}\Delta^{1/2}}_6^6.
    \end{split}
    \label{eq:A2}
\end{align}
The first line of (A2) may be written in coordinate space.
Then  the second line follows from Kato's inequality in the form \cite{12,13,14,33,41,42,43,44,45,46}
\begin{align}
    \abs{\Delta_A(x,y)}\le \Delta(x-y),
    \label{eq:A3}
\end{align}
where    $\Delta(x) = mK_1(mx)/(4\pi^2x)$, and $K_1$  is a modified Bessel  function.
We also made use of the identity 
\begin{align}
    \FD(x,y) = \frac{1}{\pi} \int_{0}^{\infty}\frac{\mathrm da}{\sqrt a}
    \bra{x} \frac{1}{(P-A)^2+m^2+a}\ket{y},
    \label{eq:A4}
\end{align}
to obtain $\abs{\FD(x,y)} < \Delta^{1/2}(x-y)$ from (A3) with $\Delta^{1/2}(x)=(m/(2\pi^{5/3}x))^{3/2}K_{3/2}(mx)$.
This result for $\Delta^{1/2}$ is obtained from  representation (A4) with $A_\mu=0$ using integral 2.16.3.8 of \cite{35}.

To prove that $\abs F^{1/2}\Delta^{1/2}\in\mathscr I_6$ it has to be shown that this  operator maps $L^2 (\R4)$ into $L^2(\R4)$ for $F_{\mu\nu}\in\cap_{q>2}L^2(\R4)$.
Let $\varphi=\abs F^{1/2}\FD\psi$, $\psi\in L^2$.
Then by Kato's inequality
\begin{align}
    \begin{split}
        \norm\varphi_2^2&=\int\psi^*\FD\abs F\FD\psi\\
        &\le \int\abs\psi\Delta^{1/2}\abs F\Delta^{1/2}\abs\psi.
    \end{split}
    \label{eq:A5}
\end{align}
Let   $\rho(x)= \int\mathrm d^4y\, \Delta^{1/2}(x-y)\abs{\psi(y)} = \Delta^{1/2}\star\abs\psi(x)$.
By H\"older's  inequality
\begin{align}
    \norm\varphi_2\le\norm*{\abs F^{1/2}\rho}_2\le\norm\rho_p\norm*{\abs F^{1/2}}_q,
    \label{eq:A6}
\end{align}
where $1/p + 1/q = 1/2$, $p$, $q\ge1$.
Since we assume $q>4$ in (A6) then  $1\le p<4$.
Use Young's inequality in the form given in Table IX.1 of \cite{40}, 
\begin{align}
    \norm{f\star g}_r\le\norm f_s\norm g_t,
    \label{eq:A7}
\end{align}
with $1/s + 1/t = 1 + 1/r$, $1\le r$, $s$, $t\le\infty$.
Then $\norm\rho_p=\norm{\Delta^{1/2}\star\abs\psi}_p\le\norm{\Delta^{1/2}}_r\norm\psi_2$, $r<4/3$.
As $\Delta^{1/2}(x)$ behaves  as $1/x^3$  for $x\to0$ and exponentially decreases for $x\to\infty$  then $\norm{\Delta^{1/2}}_r<\infty$, proving that $\varphi\in L^2$.

To complete the proof that $\abs F^{1/2}\Delta^{1/2}\in\mathscr I_6$ we rely on the  following theorem specialized to four dimensions \cite{8,47}.

 Theorem A: Let $f(x)g(-i\nabla)$ map $L^2(\R4)$ into $L^2(\R4)$.

If $f\in L^r(\mathrm d^4x)$  and $g\in L ^r(\mathrm d^4 p)$ with $2\le r<\infty$, then $f(x)g(-i\nabla )$ is in     $\mathscr I_r$ and
\begin{align}
    \norm{f(x)g(-i\nabla)}_{\mathscr I_r}\le
    (2\pi)^{-4/r}\norm f_{L^r}\norm g_{L^r}.
    \label{eq:A8}
\end{align}
We have just shown that $\abs F^{1/2}\Delta^{1/2}$ is a bounded operator on $L^2( \R4)$, for  $F_{\mu\nu}\in\cap_{q>2}L^q(\R4  )$.
By inspection $\abs F^{1/2}\in L^6(\mathrm d^4x)$ and $(p^2+m^2 )^{-1/2}\in   L^6 (\mathrm d^4p  )$  and hence $\abs F^{1/2}\Delta^{1/2}\in\mathscr I_6$.
This establishes that $\FD\abs F\FD\in\I3$ on referring  to (A1) and (A2), and hence so does $\FD\sigma F\FD$.

      Finally, in both Sec.VB and Appendix D it is claimed that  if $\varphi\in L^2$ then so does $\psi=\FD\varphi$.
We have
\begin{align}
    \abs{\psi(x)} &\le \int\mathrm d^4\,\abs{\FD(x,y)}\abs{\varphi(y)}\\
    &\le \int\mathrm d^4y\, \Delta^{1/2}(x-y)\abs{\varphi(y)} = \Delta^{1/2}\star\abs\varphi(x).
    \label{eq:A9}
\end{align}
Then by Young's inequality (A7), $\norm\psi_2\le\norm{\Delta^{1/2}\star|\varphi|}_2\le\norm{\Delta^{1/2}}_1\norm\varphi_2<\infty$ since $\norm{\Delta^{1/2}}_1<\infty$.

\newpage
\begin{widetext}
\section{Equivalence of the two sides of Eq. \eqref{eq:threepointsix}}
Reduce notation by setting $B= \frac{1}{2} \sigma \text{F}$ and $e=1$.
Begin with the  right-hand side of (3.6) by substituting (3.5) and obtain
  \begin{align}
    \begin{split}
      \text{RHS} = &\int_{0}^{\infty} \frac{\text{d}t}{t} e^{-tm^2} \text{Tr} \Big( e^{-tD^2} - e^{-t(D^2 + B)} - \int_0^{t} \text{d}t e^{-(t-s)D^2}Be^{-sD^2} \\
	    &+ \int_0^t \text{d}s_1 \int_0^{t-s_1}\text{d}s_2 e^{-(t-s_1-s_2)D^2} B e^{-s_2 D^2} B e^{-s_1 D^2} \Big).
    \end{split}
    \label{eq:B1}
  \end{align} 
Eliminate the O($B$) term by taking the spin trace of this term.
Then
  \begin{align}
    \begin{split}
      \frac{\text{d}\text{(RHS)}}{\text{d}m^2} = &\text{Tr}  \left[ \left( D^2+B+m^2 \right)^{-1} - \left( D^2 + m^2 \right)^{-1} \right] \\
	    &- \int_0^{\infty} \text{d}t e^{-tm^2} \text{Tr} \left[ \int_0^t \text{d}s_1 \int_0^{t-s_1} \text{d}s_2 e^{-(t-s_1-s_2)D^2}Be^{-s_2D^2}Be^{-s_1D^2} \right].
    \end{split}
    \label{eq:B2}
  \end{align} 
Note that
  \begin{align}
    \begin{split}
      \left( D^2 + B +m^2 \right)^{-1} -\left( D^2 + m^2 \right)^{-1} = &-\frac{1}{D^2+m^2}B\frac{1}{D^2+m^2} + \frac{1}{D^2+m^2}B\frac{1}{D^2+m^2}B\frac{1}{D^2+m^2}\\
		    &- \frac{1}{D^2+B+m^2}B\frac{1}{D^2+m^2}B\frac{1}{D^2+m^2}B\frac{1}{D^2+m^2}.
    \end{split}
    \label{eq:B3}
  \end{align} 
Substitute \eqref{eq:B3} in \eqref{eq:B2} and eliminate the O($B$) term by tracing  over its spin to get 
  \begin{align}
    \begin{split}
      \frac{\text{d}\text{(RHS)}}{\text{d}m^2} =& \text{Tr} (R) + \text{Tr} \left( \Delta_A B \Delta_A B \Delta_A \right) \\
		      & - \int_0^{\infty} \text{d} t e^{-tm^2} \text{Tr} \left[ \int_0^t \text{d}s_1 \int_0^{t-s_1} \text{d}s_2 e^{-(t-s_1-s_2)D^2}Be^{-s_2D^2}Be^{-s_1D^2}\right],
    \end{split}
    \label{eq:B4}
  \end{align} 
where 
  \begin{align}
    R = - \frac{1}{D^2+B+m^2} B \Delta_A B \Delta_A B \Delta_A.
    \label{eq:B5}
  \end{align}
  
The trace of R is obviously finite.
The second trace in  \eqref{eq:B4} is cancelled by the last integral.
To see this use the  cyclic property of the trace in the last integral and integrate the $s_1$-integral by  parts to obtain
  \begin{align}
    \begin{split}
      \frac{\text{d}\text{(RHS)}}{\text{d}m^2} =& \text{Tr} (R) + \text{Tr} \left( \Delta_A B \Delta_A B \Delta_A \right) \\
		      & - \int_0^{\infty} \text{d} t  \text{Tr} \left( e^{-(D^2+m^2)t} \int_0^t \text{d} s se^{sD^2}Be^{-sD^2}B \right).
    \end{split}
    \label{eq:B6}
  \end{align} 
  The trace manipulations here and below are allowed due to the  presence of the exponentiated (bounded) operators.
Now integrate the  $t$-integral by parts twice, firstly to get rid of the $s$-integration,  and secondly to eliminate the factor $t$ to obtain     
  \begin{align}
    \begin{split}
      \frac{\text{d}\text{(RHS)}}{\text{d}m^2} =& \text{Tr} (R) \\
		      =& -\frac{1}{8} \text{Tr}\left( \frac{1}{D^2 + \frac{1}{2}\sigma F + m^2} \sigma F \Delta_A \sigma F \Delta_A \sigma F \Delta_A \right).
    \end{split}
    \label{eq:B7}
  \end{align} 
 Now relate the left-hand side of \eqref{eq:threepointsix} to the result \eqref{eq:B7}.
We  know that $T \equiv \Delta_A^{1/2}\frac{1}{2} \sigma F \Delta_A^{1/2} \in \mathscr I_3$.
Then \cite{11} 
  \begin{align}
    R_3(T) \equiv (1+T)e^{-T+T^2/2} - 1 \in \mathscr I_1 ,
    \label{eq:B8}
  \end{align} 
so that the the relation $\text{lndet}(1+R_3) = \text{Trln}(1+R_3)$ is valid.
From  the definition \eqref{eq:threepointseven} this gives 
  \begin{align}
    \text{lndet}_3(1+T) = \text{Tr}\left[ \ln{(1+T)} -T + \frac{1}{2} T^2 \right].
    \label{eq:B9}
  \end{align} 
Noting that
  \begin{align}
    \frac{\text{d}T}{\text{d}m^2} = -\frac{1}{2} \Delta_A T - \frac{1}{2}T \Delta_A,
    \label{eq:B10}
  \end{align} 
differentiation of \eqref{eq:B9} with respect to $m^2$ gives
  \begin{align}
    \begin{split}
      \frac{\text{d}}{\text{d}m^2}\text{lndet}&_3\left(1+ \Delta_A^{1/2}\frac{1}{2} \sigma F \Delta_A^{1/2}\right)   = -\text{Tr} \left(\Delta_A \frac{1}{1+T}T^3 \right)\\ 
		      =& -\frac{1}{8} \text{Tr}\left( \frac{1}{D^2 + \frac{1}{2}\sigma F + m^2} \sigma F \frac{1}{D^2+m^2}\sigma F \frac{1}{D^2+m^2}\sigma F \frac{1}{D^2+m^2} \right)\\
		      =& \frac{\text{d}\text{(RHS)}}{\text{d}m^2}.
    \end{split}
    \label{eq:B11}
  \end{align} 
Since both sides of \eqref{eq:threepointsix} vanish for $m = \infty$ then the two sides are equivalent on integrating \eqref{eq:B11}. 
\end{widetext}

\section{Simplification of Eq. (\ref{eq:threepointeight})}

    Refer to the last term in (3.8) and take the spin trace.
Denoting  this term by $\Pi$ it is
\begin{widetext}
    \begin{align}
        \Pi = e^2\int_{0}^{\infty}\frac{\mathrm dt}{t}\,e^{-tm^2}
        \left[ \frac{\norm{F}^2}{32\pi^2}-\TR\int_{0}^{t}\mathrm ds_1\,\int_0^{t-s_1}\mathrm ds_2\,
            e^{-(t-s_1-s_2)D^2}
            F_{\mu\nu}e^{-s_2D^2}F_{\mu\nu}e^{-s_1D^2}
        \right].
        \label{eq:C1}
    \end{align}
To $\mathrm O(e^2)$ (C1) gives 
    \begin{align}
        \Pi = \frac{e^2}{32\pi^2} \int_{0}^{1}\mathrm dz\,\int\frac{\mathrm d^4k}{(2\pi)^4}
        \abs{\hat F_{\mu\nu}(k)}^2
        \ln\left( \frac{k^2z(1-z)+m^2}{m^2} \right)
        + \mathrm O(e^4),
        \label{eq:C2}
    \end{align}
verifying that   $\Pi$ is finite and that   $\Pi(m= \infty)=0$, as inspection of  (C1) indicates.

To simplify \eqref{eq:C1} integrate  the $s_1$-integral by parts, use the cyclic property of the trace,  and let $s_1= s$ to get
    \begin{align}
        \Pi = e^2\int_{0}^{\infty}\frac{\mathrm dt}{t}\,e^{-tm^2}
        \left[ \frac{\norm{F}^2}{32\pi^2}-\TR\int_{0}^{t}\mathrm ds\,s
            e^{-(t-s)D^2}
            F_{\mu\nu}e^{-sD^2}F_{\mu\nu}
        \right].
        \label{eq:C3}
    \end{align}
It is safe to differentiate $\Pi$ with respect to $m^2$  as this makes (C3)  even more ultraviolet convergent.
Doing this and integrating the  $t$-integral by parts gives    
\begin{align}
    \begin{split}
        -\frac{\mathrm d\Pi}{\mathrm dm^2}
        &= e^2 \int_{0}^{\infty}\mathrm dt \, e^{-tm^2}
        \left[ \frac{\norm F^2}{32\pi^2} - t\TR \left( \frac{1}{D^2+m^2}F_{\mu\nu}e^{-tD^2}F_{\mu\nu} \right) \right]\\
        &= e^2 \int_{0}^{\infty}\mathrm dt \, 
        \left[  e^{-tm^2}\frac{\norm F^2}{32\pi^2}
            + \int_{0}^{\infty}\mathrm ds\, \TR \left( e^{-s(D^2+m^2)}F_{\mu\nu}\frac{\mathrm d}{\mathrm dm^2}
        e^{-t(D^2+m^2)}F_{\mu\nu}\right)\right]\\
        &= e^2\frac{\mathrm d}{\mathrm dm^2}\int_{0}^{\infty}\mathrm dt\,
        \left[ - e^{-tm^2}\frac{\norm F^2}{32\pi^2t}
            +\frac{1}{2} \int_{0}^{\infty}\mathrm ds\, \TR \left( e^{-s(D^2+m^2)}F_{\mu\nu}e^{-t(D^2+m^2)}F_{\mu\nu}\right)
        \right]   . 
    \end{split}
    \label{eq:C4}
\end{align}
Hence, 
\begin{align}
    \Pi = e^2 \int_{0}^{\infty}\mathrm dt\, e^{-tm^2}
        \left[  \frac{\norm F^2}{32\pi^2t}
            -\frac{1}{2} \TR \left( e^{-tD^2}F_{\mu\nu}\Delta_AF_{\mu\nu}\right)
        \right]  ,
    \label{eq:C5}
\end{align}
since  $\Pi (m=\infty)=0$.
This is the result in (3.9).

      As a check on (C5), its $\mathrm O(e^2)$ expansion reproduces the result  (C2).
      In (3.9) $\det_3$  has no $\mathrm O(e^2)$ term by its definition, and  $\ln\det_\mathrm{SQED}$ in (3.3) to $\mathrm O(e^2)$ is 
      \begin{align}
          \ln\mbox{$\det_\mathrm{SQED}$} =
          - \frac{e^2}{64\pi^2}\int_{0}^{1}\mathrm dz \,
          (1-2z)^2 \int\frac{\mathrm d^4k}{(2\pi)^4}\,
          \abs{\hat F_{\mu\nu}(k)}^2 \ln \left( \frac{k^2z(1-z)+m^2}{m^2} \right)
          + \mathrm O(e^4).
          \label{eq:C6}
      \end{align}
Combining (C2) with (C6) following (3.9) gives the textbook result  for the lowest-order vacuum polarization graph with on-shell renormalization:
      \begin{align}
          \ln\mbox{$\det_\mathrm{ren}$} =
          \frac{e^2}{8\pi^2}\int\frac{\mathrm d^4k}{(2\pi)^4}\,
            \abs{\hat F_{\mu\nu}(k)}^2 
          \int_{0}^{1}\mathrm dz \,
          z(1-z) 
          \ln \left( \frac{k^2z(1-z)+m^2}{m^2} \right)
          + \mathrm O(e^4).
          \label{eq:C7}
      \end{align}
\end{widetext}

\section{Eigenvalue pairs of $\Delta_A^{1/2} \sigma F \Delta_A^{1/2}$}
From the equation for the scalar field propagator in the  external potential $A_{\mu}$,
  \begin{align}
    \left[ \left( \frac{1}{i} \partial_{\mu} - e A_{\mu} \right)^2 + m^2 \right] \Delta_A(x,y) = \delta(x-y),
    \label{eq:D1}
  \end{align} 
obtain by inspection
  \begin{align}
    \Delta_{A+\partial \lambda}(x,y) = e^{i e (\lambda(x)-\lambda(y))}\Delta_A(x,y).
    \label{eq:D2}
  \end{align}
Referring to the representation \eqref{eq:A4} of $\Delta_A^{1/2}$ conclude that it  transforms under $A \rightarrow A + \partial \lambda $ in the same way as $\Delta_A$.
Therefore, it  is evident that $\text{det}_3(1+\Delta_A^{1/2}\frac{e}{2}\sigma F \Delta_A^{1/2})$ is gauge invariant.

Noting \eqref{eq:D2}, define the gauge invariant propagator
  \begin{align}
    \tilde{\Delta}_{A}(x,y) = e^{-i e \int_y^x \text{d} \xi^{\mu} A_{\mu}(\xi)}\Delta_A(x,y).
    \label{eq:D3}
  \end{align}
In what follows it is not necessary to specify the line integral's  path.
Taking the complex conjugate of \eqref{eq:D1} deduce that $\Delta_A^* = \Delta_{-A}$ and hence from \eqref{eq:D3} that  
  \begin{align}
    \tilde{\Delta}^*_{A}(x,y) = \tilde{\Delta}_{-A}(x,y).
    \label{eq:D4}
  \end{align}
  
Refer to \eqref{eq:5.seven} and consider an eigenstate $\varphi$ of $\frac{e}{2} \Delta_A^{1/2} \sigma F \Delta_A^{1/2}$ with eigenvalue $-\lambda$.
Let $\psi = \Delta_A^{1/2}\varphi$.
Then                            
  \begin{align}
    \frac{e}{2} \Delta_A\sigma F \psi = -\lambda \psi.
    \label{eq:D5}
  \end{align}
Since   $\varphi \in \text{L}^2$ so does $\psi$ as shown at the end of Appendix A.
We will now show that there is an eigenstate $\psi_C$ with eigenvalue $\lambda$.

Substitute \eqref{eq:D3} in \eqref{eq:D5}:
\begin{widetext}
  \begin{align}
    \begin{split}
      \frac{e}{2}\int \text{d}^4y\tilde{\Delta}_A(x,y)\sigma F(y) &e^{-i e \int_z^y \text{d}\xi^{\mu} A_{\mu}(\xi)} \psi(y)\\
	    &= - \lambda e^{-i e \int_z^{x} \text{d}\xi^{\mu}A_{\mu}(\xi)}\psi(x),
    \end{split}
    \label{eq:D6}
  \end{align} 
\end{widetext}
where $z$ is an arbitrary point in $\mathbb{R}^4$  .
On taking the complex conjugate  of \eqref{eq:D6} we seek a matrix $C$ such that  $C \gamma_{\mu}^* C^{-1}  = -\gamma_{\mu}$.
In the  representation
\newpage
\begin{widetext}
  \begin{align}
    \boldsymbol{\gamma} = \begin{pmatrix}
0 & \boldsymbol{\sigma}  \\
\boldsymbol{-\sigma} & 0 
\end{pmatrix}, \hspace{0.2cm}     
\boldsymbol{\gamma}_0 = -i \begin{pmatrix}
0 & \mathbf{1} _{2}  \\
\mathbf{1} _{2} & 0 
\end{pmatrix}, \hspace{0.2cm}     
\boldsymbol{\gamma}_5 = \begin{pmatrix}
\mathbf{1} _{2} & 0  \\
0 & \mathbf{-1} _{2},  
\end{pmatrix},
    \label{eq:D7}
  \end{align}
\end{widetext}
one may choose $C = \gamma_3 \gamma_1$.
Since $\sigma_{\mu \nu} = [ \gamma_{\mu} , \gamma_{\nu}  ]/(2i)$, $C \sigma^*  C^{-1}  = -\sigma  $.
 Substitution of this result into the complex conjugate of \eqref{eq:D6} gives together with \eqref{eq:D4},
 \newpage
 \begin{widetext}
  \begin{align}
\frac{e}{2}\int \text{d}^4y \tilde{\Delta}_{-A}(x,y)\sigma F(y) e^{ie \int_z^y \text{d}\xi^{\mu}A_{\mu}}C\psi^*(y) = + \lambda e^{ie \int_z^x \text{d}\xi^{\mu}A_{\mu}}C\psi^*(x).
    \label{eq:D8}
  \end{align}
\end{widetext}

 $\tilde{\Delta}_A$ is gauge invariant and depends only on $F_{\mu \nu}$  and invariants  derived from it.
The second line in \eqref{eq:5.2} when expanded in powers  of $T$ consists of loops with $\tilde{\Delta}_A$ between insertions of $\sigma F$ as the phase  factors from $\Delta_A$ cancel in the trace.
Since $\text{lndet}_3$ is real, $\tilde{\Delta}_A$ is real and hence by \eqref{eq:D4} $\tilde{\Delta}_{-A} = \tilde{\Delta}_A$, expressing  $C$-invariance.
Inserting this result in \eqref{eq:D8} we conclude that for each eigenstate $\psi$ of $\frac{e}{2}\Delta_A \sigma F$ with eigenvalue $-\lambda$   there is a paired  eigenstate         
   \begin{align}
    \psi_C (x) = e^{2 i e \int_z^x \text{d}\xi^{\mu} A_{\mu}(\xi)} C \psi^*(x),
    \label{eq:D9}
  \end{align}
 
 with eigenvalue $+\lambda$.

\section{Calculation of $\lambda$}
Substitute either of the expansions \eqref{eq:5.46}, \eqref{eq:5.47} or \eqref{eq:5.48},  \eqref{eq:5.49} to O($m^2$) in \eqref{eq:5.42} and obtain using \eqref{eq:5.50}       
  \begin{align}
f_2'' + \left( \frac{4j + 3}{r} - 2 era \right)f_2' = 1- 4ea \delta_2 - er \frac{\text{d}a}{\text{d}r} \delta_2.
    \label{eq:E1}
  \end{align}
The solution of \eqref{eq:E1} at $r=R$ that is finite at $r=0$ is
\begin{widetext}
  \begin{align}
f_2'(R) = \int_0^R \text{d} r \left( \frac{r}{R} \right)^{4j+3}\left( 1 - 4e \delta_2 a(r) - e \delta_2 r a'(r) \right) e^{2e \int_r^R\text{d}s s a(s)}.
    \label{eq:E2}
  \end{align}
\end{widetext}
To $O(m^2$) the boundary condition \eqref{eq:5.45} requires
  \begin{align}
R f_2'(R) = \frac{R^2/2}{2j+2-e\nu} + \frac{e\nu \delta_2}{e\nu-2j-1}.
    \label{eq:E3}
  \end{align}
Note that $a(r)$, regardless of the sign of $C$ in \eqref{eq:5.40}, approaches   $\nu /r^2$  as $r \nearrow R$.
Therefore, $f_2'(R)$ in \eqref{eq:E2} is exponentially increasing with $e$  while the right-hand side of \eqref{eq:E3} has no such exponential growth.
 Accordingly, the boundary condition \eqref{eq:E3} requires $\delta_2$ to satisfy
  \begin{align}
    \delta_2 = \frac{\int_0^R \text{d}r \left( \frac{r}{R} \right)^{4j+3} e^{2e\int_r^R\text{d}s sa}}{e \int_0^R \text{d}r \left( \frac{r}{R} \right)^{4j+3}\left( 4a + ra' \right) e^{2e\int_r^R \text{d} s s a}} + c,
    \label{eq:E4}
  \end{align}
where $c$ is an exponentially decaying function of $e$.
Insert \eqref{eq:E2} in  \eqref{eq:E3} and then refer to \eqref{eq:E4} to obtain an equation for $c$:
\begin{widetext}
  \begin{align}
    ceR \int_0^R \text{d} r \left( \frac{r}{R} \right)^{4j+3} \left( 4a + ra' \right) e^{2e \int_r^R \text{d}s sa} = \frac{R^2/2}{e\nu-2j-2} + \frac{e\nu\delta_2}{2j+1-e\nu}. 
    \label{eq:E5}
  \end{align}
\end{widetext}
As $\delta_2$ is determined by \eqref{eq:E4} up to an exponentially decaying term, \eqref{eq:E5} is sufficient to determine $c$.

It remains to estimate $\delta_2$  in \eqref{eq:E4} with $e\nu > 2j+2$ and $e\gg1$.
 The structure of the first term in \eqref{eq:E4} suggests Laplace's  method \cite{16} as the most direct way of proceeding.
Consider the  numerator of \eqref{eq:E4}:
  \begin{align}
    I = \frac{1}{R} \int_0^R \text{d}r \left(\frac{r}{R}\right)^{4j+3}e^{2e\int_r^R \text{d}ssa}.
    \label{eq:E6}
  \end{align}
Let $r=xR$, $s=tR$ and set
  \begin{align}
    g(x) = (4j+3) \ln(x) + 2eR^2\int_x^1 \text{d}t ta.
    \label{eq:E7}
  \end{align}
Let $g'(x_0) = 0$.
Since $e \nu > 2j+2$, $g'(1)<0$ and $g'(x) \rightarrow \infty$   for $x \searrow 0$ then $g''(x_0)<0.$
 Hence, $0<x_0 <1$.
For any sign of $C$ in \eqref{eq:5.40} and $\epsilon \ge 2$ a sketch of $(4j+3)/x$ and $2eR^2xa$ versus $x$ indicates that  $4a(x_0)+x_0 a'(x_0 ) > 0$.
These strong statements can be made due to  the simplicity of $a$ in \eqref{eq:5.40}.
Therefore, for $e\gg1$
  \begin{align}
    I = e^{g(x_0)} \sqrt{\frac{2\pi}{\left| g''(x_0) \right|}}\left( 1 + O(g^{i\nu}(x_0)/e^2) \right). 
    \label{eq:E8}
  \end{align}
Since $a(r)$ is a smooth function for $0<r<R$, $g^{i\nu} (x_0 )$ is finite and $O(e)$ or less.
Repeating this procedure for the denominator of \eqref{eq:E4}  gives for $e\nu  >2j+2$, $e\gg1$
  \begin{align}
    \delta_2 = \frac{1/e}{4a(r_0)+r_0 a'(r_0)}\left( 1 + O(1/e) \right) > 0,
    \label{eq:E9}
  \end{align}
where $r_0 =Rx_0$  is the unique root in the interval $0<r<R$ of
  \begin{align}
    4j+3-2er^2a(r) = 0.
    \label{eq:E10}
  \end{align}
   Refer to \eqref{eq:5.14} and define the spin trace norm of an operator $A$ by $\norm{A}_1 = \text{Tr}\left( A^{\dagger} A\right)^{1/2}$ so that $\frac{1}{2} \norm{(\sigma F)^+}_1 = \left| 4a+ra' \right|$,
   where $(\sigma F)^+$ is defined by \eqref{eq:5.14}.
 Then \eqref{eq:E9} becomes
   \begin{align}
    \delta_2 = \frac{2}{e\norm{(\sigma F(r_0))^+}_1}\left( 1 + O(1/e) \right).
    \label{eq:E11}
  \end{align}
Here $F_{\mu \nu} (r_0)$ is a smoothly varying function on $0<r_0 <R$ and hence  slowly varying for $j = 0,1/2 ,.,j_{max}$ and $e\nu  > 2j+2$, $e\gg1$.

Repeated application of Laplace's method gives the following  additional results for $e\gg1$.
For $j = 0,1/2 ,.,j_{max} -1/2$, $e\nu  > 2j+2$ with  $e\nu  = N+\Delta$  , $0< \Delta <1$, $N=2,3,.,j_{max} = (N-2)/2$, $\delta_4$  in \eqref{eq:5.47} is   
   \begin{align}
    \delta_4 = -\delta_2^2 + O\left( \frac{R^4}{e^4} \right).
    \label{eq:E12}
  \end{align}
For $j = j_{max}$, $\delta_{2\alpha_0}$ in \eqref{eq:5.49} exponentially decreases with $e$ and the  $O(m^4 )$ term is the same as that in \eqref{eq:5.47} with $\delta_4$ given by \eqref{eq:E12}.
 For $e\nu  = 3,4...$ and $j=j_{max} = (N-3)/2$, \eqref{eq:5.47} holds with $\delta_2$, $\delta_4$ given  by \eqref{eq:E11} and \eqref{eq:E12}.

\section{Zero mass limit of $\boldsymbol{\det_\mathrm{{ren}}}$}

The renormalized determinant in \eqref{eq:twopointone} may be equivalently  expressed as \cite{7,31,48}
  \begin{align}
      \text{det}_{ren} \left( 1 - eS \slashed{A} \right) = \exp \left(\Pi_2+\Pi_3+\Pi_4\right)\text{det}_5\left( 1- eS\slashed{A} \right),
    \label{eq:F1}
  \end{align} 
where
  \begin{align}
      \text{lndet}_{5} \left( 1 - eS \slashed{A} \right) = \text{Tr} \left[ \ln \left( 1 - eS \slashed{A} \right) + \sum_{n=1}^4 \left(eS \slashed{A} \right)^n/n \right].
    \label{eq:F2}
  \end{align} 
As evident from \eqref{eq:F1}, det$_5$ is the remainder of det$(1-eS\slashed A)$ after the $O(e^2 ,e^3 ,e^4  )$ graphs $ \Pi_2, \Pi_3$ and $\Pi_4$ have been factored out.
To maintain equality with \eqref{eq:twopointone} they are defined by the power series expansion of its right-hand side to $O(e^4)$.
This definition  gives the on-shell subtracted vacuum polarization graph $\Pi_2$ in \eqref{eq:C7}; $\Pi_3 = 0$, and the gauge invariant photon-photon scattering  graph $\Pi_4$.
A Hilbert space can be found on which $S\slashed A$ is a compact operator belonging to  $\mathscr I_r$ , $r>4$ provided  $A_{\mu}\in \underset{r\ge 4+\epsilon}{\cap} L^{r}$ \cite{7,31,48}.
The trace ideal  $\mathscr I_{r}$  is discussed in Sec.III and Appendix A.
Then $S\slashed A\in \mathscr I_5$ since $\mathscr I_{4+\epsilon} \subset \mathscr I_5 $, and hence det$_5$ is an entire function of $e$ of order $4$ \cite{14}.
It has no zeros for real $e$, and since det$_{\text{ren}}(e=0) = 1$, det$_{\text{ren}}   > 0$  for all real $e$.
It will now be shown that the $m=0$ limit of det$_{\text{ren}}$  is finite when $\Pi_2$ is subtracted off-shell, provided $A_{\mu} \in   \underset{r\ge 4 -\epsilon}{\cap} L^{r} ( \mathbb{R}^4 )$, $ \epsilon > 0$.
This excludes zero-mode supporting potentials that fall  off as $1/x$ and which induce divergent mass singularities in  $\ln\det_\mathrm{{ren}}$ \cite{24,49,50}.
Our analysis of the $m=0$ limit of det$_{\text{ren}}$ is a  generalization of that in \cite{31} for massless QED$_2$.\\

Instead of dealing with the operator $S\slashed A$ at $m=0$ we make a similarity transformation that leaves det$_5$ invariant. 
Setting $m=0$ let
  \begin{align}
      S\slashed A \rightarrow \frac{\slashed p}{\left| p \right|} \frac{1}{\left|p\right|^{1/2}} \left| A \right|^{1/2} \frac{\slashed A}{\left| A \right|}\left| A \right|^{1/2}\frac{1}{\left| p \right|^{1/2}},
    \label{eq:F3}
  \end{align} 
 where $\left|A\right| = (A_{\mu}^2)^{1/2}$ .
Because $\slashed p/\left|p \right|$ and $\slashed A/\left|A \right|$ are unitary it suffices to consider the operator $K = \left|p\right|^{-1/2} \left|A\right| \left|p\right|^{-1/2}$.
We claim that  $K\in \mathscr I_{r}$, $r>4$ provided $A_{\mu} \in \underset{q\ge 4-\epsilon}{\cap}L^q ( \mathbb{R}^4 ),  \epsilon >0$.
If $K \in \mathscr I_r$ then by H\"older's  inequality for $\mathscr I_r$ spaces 
\begin{align}
    \norm{K}_r \le \norm{\frac{1}{\left| p \right|^{1/2}} \left| A \right|^{1/2}}_s \norm{ \left| A \right|^{1/2}\frac{1}{\left| p \right|^{1/2}}  }_s,
  \label{eq:F4}
\end{align}
with $s=2r>8$.
If $\left|A\right|^{1/2} \left|p\right|^{-1/2} \in \mathscr I_s$ then so does its adjoint $\left|p\right|^{-1/2} \left|A\right|^{1/2}$  by  the general properties of $\mathscr I_p$ spaces.
Let
\begin{align}
    B = \left| A \right|^{1/2} \frac{1}{\left| p \right|^{1/2}} = B_1 + B_2,
  \label{eq:F5}
\end{align}
where
\begin{align}
    B_1 = \left| A \right|^{1/2} \left( \frac{1}{\left| p \right|^{1/2}} - \frac{1}{(p^2 + \mu^2)^{1/4}} \right), 
   \label{eq:F6}\\
    B_2 = \left| A \right|^{1/2} \frac{1}{(p^2 + \mu^2)^{1/4}}, 
   \label{eq:F7}
\end{align}
and where $\mu^2$ is an arbitrary mass parameter.
To prove that $B_1$, $B_2 \in \mathscr I_s$, $s>8$, it has to be first shown that these operators map $L^2 (\mathbb{R}^4)$  into $L^2 ( \mathbb{R}^4 )$.\\

We begin with $B_1$.
Let $g_1  = \Delta_1 * f$, $f\in L^2$ , where
\begin{align}
    \Delta_1(x) = \int \frac{\text{d}^4 p}{(2\pi)^4} e^{ipx}\left( \frac{1}{\left|p \right|^{1/2}} - \frac{1}{\left( p^2 + \mu^2 \right)^{1/4}} \right).
  \label{eq:F8}
\end{align}
Then $\Delta_1(x)$ behaves as $\mu^2/x^{3/2}$  for $x\rightarrow 0$ and $1/x^{7/2}$   for $x\rightarrow \infty$.
Let  $h_1  =\left|A\right|^{1/2} g_1$.
By H\"older's inequality                                                  
\begin{align}
      \norm{h_1}_2 = \norm{ \left| A\right|^{1/2} g_1 }_2 \le \norm{ \left|A\right|^{1/2} }_p \norm{g_1}_q,
    \label{eq:F9}
\end{align}
with $1/p + 1/q = 1/2$, $p,q \ge 1$.
By Young's inequality \eqref{eq:A7}, $\norm{ g_1 }_q = \norm{ \Delta_1 * f }_q \le \norm{ \Delta_1 }_r\norm{f }_2$ with $1/q + 1/2 = 1/r$, $q,r \ge 1$.
 Referring to the properties of $\Delta_1$ it is evident that $\norm{\Delta_1}_r < \infty$ provided $8/7<r<8/3$.
Choose $q > 8/3$.
From $1/p + 1/q = 1/2$ obtain $p<8$.
Then \eqref{eq:F9} allows $A_{\mu} \in \underset{p\ge 4-\epsilon}{\cap} L^p$, $\epsilon  >0$.
Under this condition $\norm{ h_1}_2 < \infty$ and hence $B_1$ is an operator on $L^2$.\\

Next consider $B_2$.
The Fourier transform of $(p^2 + \mu^2)^{-1/4}$ in  four dimensions is undefined.
So consider     
\begin{align}
    g_2(x) = \Delta_2*f(x) = \int \frac{\text{d}^4p}{(2\pi)^4} e^{ipx} \frac{\hat{f}(p)}{\left( p^2 +\mu^2 \right)^{1/4}}.
    \label{eq:F10}
\end{align}
Since  $\norm {f }_2  =  \|\hat{f}\|_2 /(2 \pi )^2$ then $\hat f(p)$ behaves as $1/p^{2+\epsilon} $   for $p\rightarrow \infty$  and $1/p^{2-\epsilon}$ for $p\rightarrow 0$.
Therefore, $g_2 (x)$ behaves as $1/x^{2+\epsilon}$   for  $x\rightarrow \infty$   and $1/x^{3/2-\epsilon}$ for $x\rightarrow 0$.
Then $B_2$ maps $L^2$ into $L^2$  since  $h_2  = \left|A\right|^{1/2} g_2$ satisfies  $\norm{ h_2 }_2    =   \norm{ \left|A \right|^{1/2} g_2 }_2 \le \norm{ \left| A\right|^{1/2}}_p \norm{g_2 }_q$ with $1/p + 1/q = 1/2$, $p,q\ge1$.
Thus $\norm{ g_2 }_q   < \infty$ for $2<q<8/3$ which requires $p>8$ or $A_{\mu} \in \underset{p>4}{\cap}L^p$.

To complete that proof that $B_1$ , $B_2 \in \mathscr I_s$, $s>8$ we rely on Theorem A  in Appendix A.
For $B_1$, since
\begin{align}
    \frac{1}{\left| p \right|^{1/2}} - \frac{1}{\left(p^2 +\mu^2 \right)^{1/4} }&= O\left( \left| p \right|^{-3/2} \right), \left|p \right| \rightarrow \infty \notag \\ 
										&= O \left( \left|p\right|^{-1/2} \right), \left|p \right| \rightarrow 0,
  \label{eq:F11}
\end{align}
the left-hand side belongs to $L^s (\text{d}^4 p)$ for $8/3<s<8$.
It was  just shown that $B_1$  is a bounded operator on $L^2$  if $\left|A\right|^{1/2} \in  L^s (\text{d}^4 x)$,  $s<8$.
By Theorem A $B_1 \in \mathscr I_s$, $s<8$, and therefore by the general  properties of $\mathscr I_p$ spaces, $B_1 \in \mathscr I_s$, $8\le s\le\infty$.
     
For $B_2$ evidently $(p^2  + \mu^2  )^{-1/4} \in L^s (\text{d}^4 p)$ for $s>8$.
For $B_2$  to be  a bounded operator on $L^2$  it was found that $\left|A\right|^{1/2} \in L^s (\text{d}^4 x), s>8$.
 Hence $B_2\in \mathscr I_s$, $8<s\le \infty$ by Theorem A.

It has now been established that $B_1+B_2 =\left|A\right|^{1/2}\left|p\right|^{-1/2}\in \mathscr I_s$, $ 8<s\le\infty$ provided $A_{\mu} \in \underset{r \ge 4 -\epsilon}{\cap}L^r $, $ \epsilon>0$. Referring to \eqref{eq:F4},  $K = \left|p\right|^{-1/2} \left|A\right| \left|p\right|^{-1/2}$ $\in \mathscr I_r, 4<r\le \infty$, and hence det$_5$  is well-defined at  $m=0$ since $K\in \mathscr I_5$.
The loop expansion of det$_5$ makes sense, and so  the similarity transformation defined in \eqref{eq:F3} is valid, allowing  us to conclude that $S\slashed A|_{m=0} \in \mathscr I_5$ for the restricted class of $A_{\mu}$ potentials considered here.

It remains to demonstrate the continuity of the $m=0$ limit of det$_5(1-eS\slashed A) = \text{det}_5 (1-e\slashed AS)$ for $m>0$.
We will deal with the operator $\slashed A S$.
The continuity of the $m=0$ limit of det$_5$ will follow from a theorem Gohberg and Kre\v{\i}n, Ch. 4, Th. 2.1 \cite{10}: Let $A \in \mathscr I_p$,  where $p$ is a positive integer, and let $F$ be an arbitrary closed  bounded set.
Then for any $\epsilon  >0$ there exists a $\delta >0$ such that for  any operator $B \in \mathscr I_p$,
\begin{align}
 \max_{\mu \in F} \left| \text{det}_p \left( 1- \mu A \right) - \text{det}_p \left( 1- \mu B \right) \right| < \epsilon \notag
\end{align}
whenever $\norm{A-B }_p < \delta   $.\\
Consider                             
\begin{align}
    \slashed A S - \slashed A S_{m=0} = \slashed A \frac{m^2 \slashed p}{p^2(p^2+ m^2)} + \slashed A \frac{m}{p^2 + m^2}. 
\end{align}
It is now known that $\slashed A S$, $\slashed A S_{m=0} \in \mathscr I_5$ for $A_{\mu} \in \underset{r\ge 4-\epsilon}{\cap}L^r$, $\epsilon  >0$.
Then
\begin{align}
   \norm{ \slashed A S - \slashed A S_{m=0} }_5 \le \norm{ \slashed A \frac{m^2 \slashed p}{p^2(p^2+m^2)} }_5 + \norm{ \slashed A \frac{m}{p^2+m^2} }_5.
  \label{eq:F13}
\end{align}
Let    
\begin{align}
    B_3 = \slashed A \frac{\slashed p}{p^2(p^2+m^2)},
   \label{eq:F14}
\end{align}
where $B_3$  is an operator on $L^2$  for $A_{\mu}$  restricted as above.
The proof  of this proceeds in exactly the same way as in the case of $B_1$ above.
The form of $B_3$ allows immediate application of Theorem A, Appendix A.
By inspection $\slashed p/[p^2 (p^2 + m^2  )]\in  L^{4-\epsilon}   (\text{d}^4 p), \epsilon  >0$, and hence  $B_3 \in \mathscr I_{4-\epsilon}$.
Letting  
\begin{align}
    B_4 = \slashed A \frac{1}{p^2+m^2},
   \label{eq:F15}
\end{align}
we conclude by the same analysis that $B_4 \in \mathscr I_{4-\epsilon}$.

      It is a general property of $\mathscr I_p$ spaces that $\norm{ T }_p   \le \norm{ T }_{p'}$, $p\ge p'$.
 Thus, from \eqref{eq:F13}, 
 \begin{align}
    \norm{ \slashed A S - \slashed A S_{m=0} }_5 \le m^2 \norm{ B_3 }_{4-\epsilon} + m \norm{ B_4 }_{4-\epsilon}.
   \label{eq:F16}
\end{align}
Referring again to \eqref{eq:A8} Theorem A obtain  
\begin{widetext}
  \begin{align}
      \norm{ \slashed A S - \slashed A S_{m=0} }_5 \le (2\pi)^{\frac{4}{4-\epsilon}}\norm{ \slashed A }_{4-\epsilon} \left( m^2\norm{\frac{\slashed p}{p^2(p^2+m^2)} }_{4-\epsilon} + m \norm{ \frac{1}{p^2+m^2} }_{4-\epsilon}\right).
    \label{eq:F17}
  \end{align} 
\end{widetext}
The two $L^{4-\epsilon}  (\text{d}^4 p)$ norms on the right-hand side of \eqref{eq:F17} multiplied  by $m^2$  and $m$ both vanish as $m^{\epsilon/(4-\epsilon)}$ as $m\rightarrow 0$ when $p$ is rescaled to m$p$.

This establishes the continuity of the $m=0$ limit of det$_5$ for any finite value of e by the Gohberg-Kre\v{\i}n theorem stated above.

Regarding $\Pi_2$ in \eqref{eq:F1}, we have already discussed off-shell renormalization in Sec.VIA.
Subtracting off-shell adds the term \eqref{eq:6.7} to lndet$_{\text{ren}}$.
When this is combined with the right-hand side of \eqref{eq:C7}, which defines $\Pi_2$, the result is $\underset{m=0}{\lim}\Pi_2=$finite.

     Finally, the $m=0$ limit of the photon-photon scattering  graph $\Pi_4$ has been considered in detail for potentials with a $1/x$ fall off \cite{51}.
The conclusion is that $\underset{m=0}{\lim} \Pi_4 =$finite.
The inclusion of potentials with a faster fall off such as those  considered here can only reinforce this conclusion.

Summarizing, it has been established that $\underset{m=0}{\lim} \text{ lndet}_{\text{ren}}  =$finite  for off-shell charge renormalization and potentials $A_{\mu} \in \underset{r\ge 4 - \epsilon}{\cap} L^r ( \mathbb{R}^4)$.
For zero mode supporting potentials the zero mass limit of lndet$_{\text{ren}}$  is not finite, but we know precisely where this divergence  occurs, namely in det$_3$.

\end{document}